\documentclass[showpacs,preprintnumbers,amsmath,amssymb]{revtex4}

\usepackage{color}
\usepackage{graphicx,verbatim}
\usepackage[hidelinks]{hyperref}
\usepackage{dcolumn}
\usepackage{float}
\usepackage{rotating}
\usepackage{amsmath}
\begin{document}

\newpage

\title{Beyond the Power Law: Uncovering Stylized Facts in Interbank Networks}
\author{Benjamin Vandermarliere}
\email{Benjamin.Vandermarliere@UGent.be}
\affiliation{Department of Physics and Astronomy,\\
			Department of General Economics, \\
 Ghent University, Belgium}

\author{Alexei Karas}
\email{a.karas@ucr.nl}
\affiliation{University College Roosevelt,\\ Utrecht University School of Economics, \\
The Netherlands}

\author{Jan Ryckebusch}
\email{Jan.Ryckebusch@UGent.be}
\affiliation{Department of Physics and Astronomy,\\
 Ghent University, Belgium}

\author{Koen Schoors}
\email{Koen.Schoors@UGent.be}
\affiliation{Department of General Economics, \\
 Ghent University, Belgium}

\date{\today}

\begin{abstract}
\noindent
We use daily data on bilateral interbank exposures and monthly bank balance sheets to study network characteristics of the Russian interbank market over Aug 1998 - Oct 2004. Specifically, we examine the distributions of (un)directed (un)weighted degree, nodal attributes (bank assets, capital and capital-to-assets ratio) and edge weights (loan size and counterparty exposure). We search for the theoretical distribution that fits the data best and report the ``best'' fit parameters.

\noindent
We observe that all studied distributions are heavy tailed. The fat tail typically contains 20\% of the data and can be mostly described well by a truncated power law. Also the power law, stretched exponential and log-normal provide reasonably good fits to the tails of the data. In most cases, however, separating the bulk and tail parts of the data is hard, so we proceed to study the full range of the events. We find that the stretched exponential and the log-normal distributions fit the full range of the data best. These conclusions are robust to 1) whether we aggregate the data over a week, month, quarter or year; 2) whether we look at the ``growth'' versus ``maturity'' phases of interbank market development; and 3) with minor exceptions, whether we look at the ``normal'' versus ``crisis'' operation periods. In line with prior research, we find that the network topology changes greatly as the interbank market moves from a ``normal'' to a ``crisis'' operation period.
\end{abstract}
\maketitle 

\section{Introduction}
\label{Introduction}


The frequency of an event follows a power law when that frequency varies as a power of some attribute of that event (e.g.~its size). Power-law distributions have been claimed to occur in an extraordinarily diverse range of phenomena from the sizes of wars, earthquakes and computer files to the numbers of papers scientists write and citations those papers receive \citep{newman2005power}. In economics and finance, power laws have been documented for income and wealth \cite{yakovenko2009colloquium}, the size of cities and firms, stock market returns, trading volume, international trade, and executive pay \citep{gabaix2008power}. 

Most relevant to this paper, the tail parts of interbank network characteristics, such as degree distribution, have been shown to follow a power law too (see ~\cite{cont2013network} for Brazil, ~\cite{AustriaAnalysis,cajueiro2008role} for Austria, \cite{kanno2014assessing} for Japan, and ~\cite{goddard2014size} for the commercial banks in the US). This ubiquituous presence of power laws has resulted in an extensive search for universal dynamics that can explain their existence (see \cite{de2006fitness,iori2007trading} for examples of such search in interbank networks).

Recently, however, Clauset~et.~al.~\cite{ClausetPL} (followed by ~\cite{stumpf2012critical}) call these findings into question. In particular, they critisize the commonly used methods for analyzing power-law data, such as least-squares fitting, which can produce inaccurate estimates of parameters for power-law distributions or provide no indication of whether the data obey a power law at all. Clauset~et.~al.~propose a statistical framework for discerning and quantifying power-law behavior in empirical data and apply that framework to twenty-four real-world data sets, each of which has been conjectured to follow a power law. For most datasets they find moderate to weak evidence in favor of power laws.

This debate about the potential of power laws to capture the underlying network dynamics is important for economic policy. For example, during the recent financial crisis, the interbank lending
market was one of the most important channels of financial contagion. The malfunctioning of the interconnectivity of the interbank lending network, caused a liquidity
drought with consequences reverberating throughout the entire economy. Since then, interbank markets research has proliferated. In those studies one wishes to uncover
the network topology of interbank markets, to understand how they function, and how they could catalyse a systemic meltdown \cite{rethinkingHaldane,meltdown}.
Current research on contagion in interbank markets often relies on a scale-free topology to simulate the interbank network ~\cite{krause,roukny2013}. This choice likely affects the outcome of conducted stress tests (as is explicitly confirmed by ~\cite{roukny2013}) and, therefore, the policy implications stemming from them. Yet the evidence supporting this choice is not ironclad. Understanding the properties of the tail is crucial to understand shock propagation in dynamic networks. The authors of Refs.~\cite{halaj2013assessing,georg2013effect}, among others, find that only a small fraction of possible network structures may spread relatively sizable contagion losses across the system, thus highlighting the non-linear nature of shock propagation effects and stressing that contagion is to a considerable extent a tail risk problem.


This paper contributes to the debate by providing a detailed analysis of the network characteristics of a real interbank network over an extended period of time. We use daily data on bilateral interbank exposures and monthly bank balance sheets to study network characteristics of the Russian interbank market over Aug 1998 - Oct 2004. Among other things, the analysis allows one to determine the theoretical distributions of connectivity among banks via interbank loans, crucial to assess efficiency and stability of the Russian interbank market. We focus on measures that represent essential input for most of the interbank contagion simulations. Specifically, we examine the distributions of (un)directed (un)weighted degree, nodal attributes (bank assets, capital and capital-to-assets ratio) and edge weights (loan size and counterparty exposure). Using the methodology of ~\cite{ClausetPL} we set up a horse race between the different theoretical distributions to find one that fits the data best. We then study the time evolution of the best-fit parameters. 

We observe that all studied distributions are heavy tailed. The fat tail typically contains 20\% of the data and can be systematically described by a truncated power law. In most cases, however, separating the bulk and tail parts of the data is hard, so we proceed to study the full range of the events. We find that the stretched exponential and the log-normal distributions fit the full range of the data best. Our conclusions turn out to be robust to whether we aggregate the data over a week, month, quarter or year. Further, we find no qualitative difference between the ``growth'' and ``maturity'' phases of interbank market development and little difference between the ``crisis'' and ``non-crisis'' periods.

Sec.~\ref{Sec:data} describes our data, defines the network measures we study, and summarizes the conclusions from previous studies of those measures. Sec.~\ref{Sec:methodology} and~\ref{Sec:illustration}, respectively, describe and illustrate the methodology we use to find the theoretical distribution that fits the data best. Sec.~\ref{Sec:results} reports the results. Sec.~\ref{Sec:conclusion} concludes.

\section{Data and Definitions}\label{Sec:data}
\subsection{Data Source}\label{Sec:source}
Mobile, a private financial information agency, provided us with two datasets for the period Aug 1998 - Oct 2004 \footnote{For more information on the data provider see its website at \url{www.mobile.ru}.}. The information in the datasets is a part of standard disclosure requirements and is supplied to the regulator on a monthly basis. The first dataset, described in ~\cite{karas2005heracles}, contains monthly bank balances for most Russian banks. From this dataset we take two variables: total assets and capital. The second dataset contains monthly reports ``On Interbank Loans and Deposits'' (official form's code 0409501) and represents a register of all interbank loans issued in the Russian market. For each loan we know its size, interest rate, issuer, receiver, reporting date and maturity date. On average, about half of the Russian banks are active on the interbank market. Consequently, the analysis of interbank network measures includes fewer banks than the analysis of balance sheet indicators.
With regard to the maturity of the loans, we discriminate between short-term and long-term loans. Short-term loans are defined as loans with a one-day or a one-week maturity. In this paper, we restrict ourselves to short-term loans which account for more than $80$\% of the transactions both in terms of the number and of the volume. The reasons for this restriction is because the data provide information about the repayment date and not about the issuance date of the loans. This makes it hard to infer the exact duration of the connection between two banks for the long-term loans.

\subsection{Network Measures}\label{Sec:measures}

\begin{table*}
\caption{The levels of aggregation and the number of banks for earlier empirical studies of interbank markets (overnight as well as credit networks).}
\begin{tabular}{ l  c  c  c  c }
\hline \hline
Country & Paper & Sample Period & Aggregation Level & $\#$ Banks \\
\hline
Mexican & \cite{martinez2014empirical} & 2005 - 2010 & day & 40 \\ 
Italy & \cite{iori2} & 1999 - 2002 & day & $\pm$ 200  \\
Italy & \cite{fricke2013distribution} & 1999 - 2010 & day/quarter & $\pm$ 200  \\
Japan & \cite{kanno2014assessing} & 2009m3 2013m3 & month  & 125 \\
Brazil & \cite{cont2013network} &  2007 - 2008 & month & 2400   \\
Austria & \cite{AustriaAnalysis} & 2000 - 2003 & month (only 10) & 900  \\
Germany & \cite{roukny2014network, craig2014interbank} & 2002q1 - 2012q3 & quarter & 2000+  \\
\hline \hline
\end{tabular}
\label{tab:sumStats}
\end{table*}

Before we start constructing an interbank network from our data, we stress that there are several different ways of how banks can interact. For example, one can have liability  \cite{kanno2014assessing,roukny2014network,craig2014interbank,cont2013network,AustriaAnalysis,martinez2014empirical} and overnight lending networks \cite{iori2,fricke2013distribution}, but one can also construct networks of the financial payment flows between banks, or in aggregated form between countries  \cite{kyriakopoulos2009network,soramaki,cook2014global,imakubo2010transaction,minoiu2013network,martinez2014empirical}. Banks are also connected by mutual cross holdings \cite{huang2013cascading}. In this work we choose to consider all loans between banks with a maturity of less than a week. Hence, we construct an interbank network which combines the overnight market with the short-term liability market. This enables us to keep our finger on the pulse of the interbank market, by not including longer and hence stickier contracts.

We use the two aforementioned datasets to construct an interbank network with banks as nodes and mutual contracts representing edges. This procedure is performed for aggregated data covering various time intervals. In every situation, we construct three versions of the interbank network, which differ in the level of detail in quantifying the edges. In the undirected version, an undirected unweighted edge is established between banks if they exchange money on the interbank market during the considered time period. For the directed versions, we discriminate between the issuer and the receiver of the interbank loan. A directed edge, which points from issuer to receiver, is established if the issuer has lent money to the receiver in the considered period. In the multidirected version, every interbank loan is represented by a directed edge pointing from issuer to receiver. This implies there can be several edges between pairs of nodes. The directed and undirected versions represent complementary views of the interbank network. The directed version captures the lending/borrowing activity by looking at the direction of the flows and therefore the contribution of each bank, whereas the undirected version merely captures the existence of interbank relations.

In this paper we focus on network characteristics that represent essential input for typical interbank contagion simulations ~\cite{GHK,NYYA,krause,BDGGE,georg2013effect,halaj2013assessing,kanno2014assessing,gofman2012efficiency}. We distinguish three types of network characteristics: nodal attributes, edge weights and various measures of a node's centrality. 

We consider three nodal attributes: bank capital, total assets and leverage. Total assets proxy for bank size. Capital measures its financial buffer. Leverage is defined as capital divided by total assets. Upon evaluating nodal attributes, we exclude banks with negative assets (which we attribute to data errors) or negative capital (banks in effective default).  In determining the distributions of edge weights and of node centrality measures we do not exclude banks with negative assets or negative capital.
Edge weights are measured separately for the directed and the multidirected network.  For the former, the edge weight equals counterparty exposure, that is, the total amount of money the issuer has lent to the receiver in the considered time period. For the multidirected network, the edge weight equals the size of the issued loan.

We consider various measures of a node's centrality. For an undirected network, we define the degree of a node as the number of edges connected to that node. It measures the number of counterparties a bank has on the interbank market. For the directed network, an in-degree (out-degree) of a node is the number of incoming (outgoing) edges; it measures the number of interbank lenders (borrowers) a bank has. For the multidirected network, an in-degree (out-degree) is the number of received (issued) loans. Finally, for each bank we define the total in-exposure (out-exposure) as the total amount of money borrowed from (lent to) the market.

The interbank credit network characteristics we consider in this paper have already been studied for a number of countries. Before sketching the main findings of those studies, we would like to draw the attention to Table~\ref{tab:sumStats} which summarizes  the sheer variety of considered aggregation periods and of the actual networks' size. Because of this variation, care needs to be taken when  comparing our results with those of other studies.

The majority of studies find numerous heavy tailed variables in the interbank network. In addition, many authors propose a power law as the best-fit candidate to (at least the tail of) the empirical distribution. In particular, Goddard et al.~\cite{goddard2014size} find that the asset size distribution of U.S. commercial banks is well described by a truncated lognormal, while the tail part is well fitted with a power law.  Cont et al.~\cite{cont2013network} obtain fair fits to the tails of the in-degree, out-degree, total degree, and exposure size distributions of the Brazilian interbank network with power laws. In the Austrian interbank market the loan size distribution is well described by a power law~\cite{AustriaAnalysis,cajueiro2008role}. That same study also examines the degree distribution of the undirected network and the in- and out-degree distribution of the directed alternative. Each of these three distributions is seemingly well described by two power laws, one for the low-degree region and one for the tail part. For Japan, Kanno et al.~\cite{kanno2014assessing} find a power-law distribution for the entire range of the counter-party exposure size distribution. Finally, Iori et al.~\cite{iori2} investigate the distributions of the in-degree, out-degree and exposure of the individual banks in the Italian market. The authors do not attempt to fit their data with some theoretical distribution. They do find, however, a structural difference in the network topology between the months of the global financial crisis of 2008 and normal operating months.

Few studies, however, consider alternate theoretical distributions as candidates to describe their data. The ones that do ~\cite{fricke2013distribution}, in fact, cast doubt on the ability of power laws to provide the best empirical fit. In what follows, we too consider an extensive list of alternative theoretical distributions and identify those that describe both the tail and the entire distribution better than the power law. Along the way, we apply a criterion which allows us to discriminate between the tail and the bulk parts of the data.

\subsection{Descriptive Statistics}\label{Sec:stats}
Different phases of the interbank market development (for example, growth versus maturity, or crisis versus non-crisis) may be guided by a different data generating mechanism. For example, Iori et al.~\cite{iori2} find a structural difference in the Italian interbank network topology before versus during the 2008 financial crisis. This finding suggests that the choice of the distribution that best fits the data may vary over time, and in particular may differ between crisis and non-crisis periods. In Sect.~\ref{Sec:results} we check whether such variation exists. In this section, we provide evidence that the Russian interbank market, indeed, went through some distinct development phases.

\begin{figure}
\begin{center}
\includegraphics[width=0.5\columnwidth, trim=0cm 0cm 0cm 0cm, clip=true]{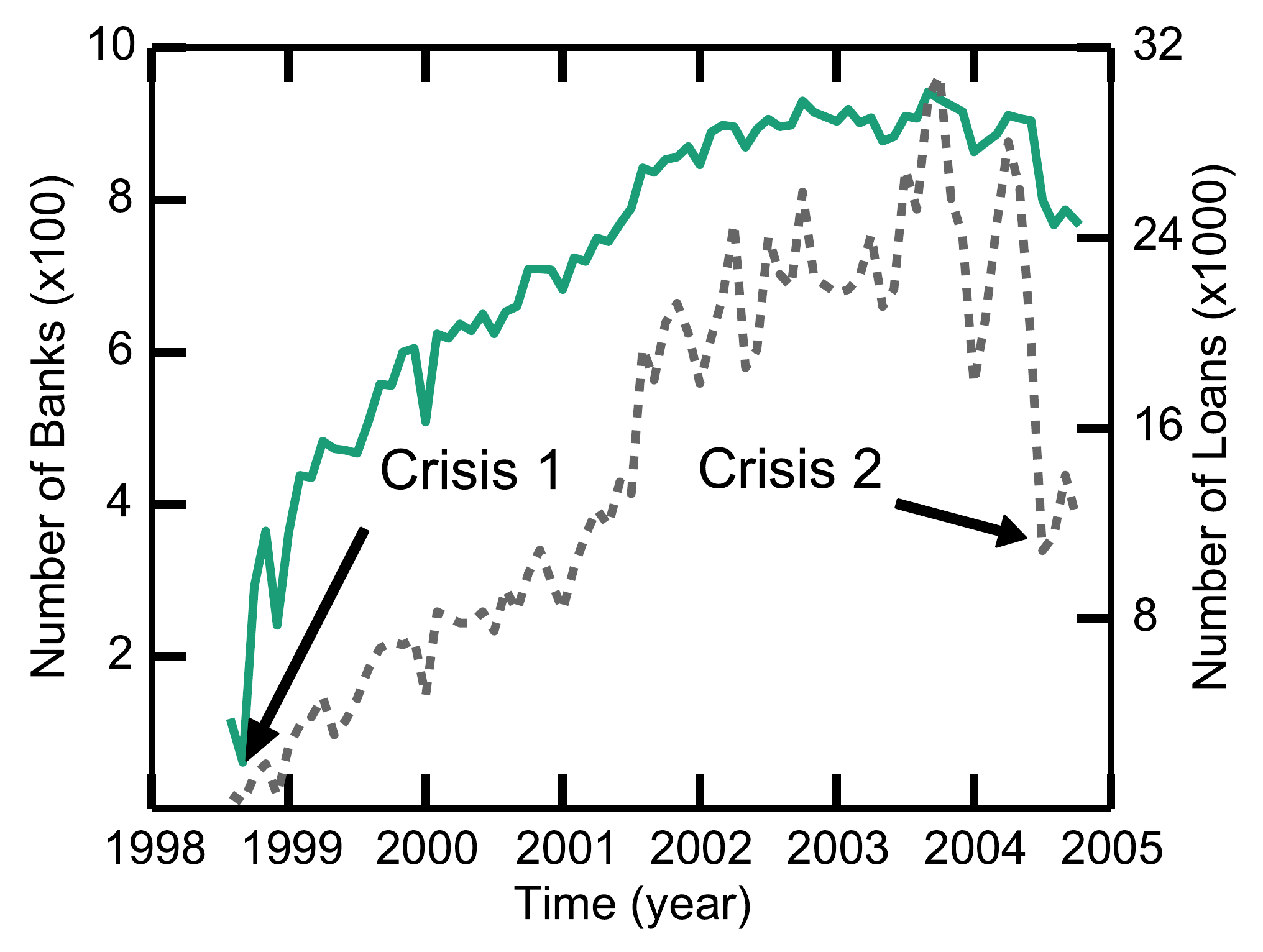}
\caption{(Color online) The time dependence of the number of active
  banks (solid line) and the number of interbank loans (dashed line)
  in the Russian interbank network between 1998 and 2005. Data
  are aggregated over a month. The arrows indicate the start of the
  two ``crises''.}
\label{fig:NodeTransactionNumber}
\end{center}
\end{figure}

Fig.~\ref{fig:NodeTransactionNumber} shows the time series of the number of banks active on the interbank market and of the number of issued loans. For this figure we use data aggregated over a month. Over 1998-2002 the interbank network experiences growth: we observe a steady and comparable increase in the number of active banks and of issued loans. After 2002, the interbank network gradually matures: the number of active banks flattens out while the number of loans per bank grows. Note, however, the strong variation in the number of issued loans from the second half of 2003 onwards.

As is clearly indicated in Fig.~\ref{fig:NodeTransactionNumber}, our sample period includes two crises: one in August 1998 and one in the summer of 2004. Both crises resulted in a partial meltdown of the Russian interbank market. They coincide with the edges of the sample period and are clearly marked by a reduction in the number of active banks and issued loans. The first crisis got triggered on August 17, 1998 when Russia abandoned its exchange rate regime, defaulted on its domestic public debt and declared a moratorium on all private foreign liabilities. The second crisis was ignited by an investigation of banks accused of money laundering and sponsorship of terrorism. This gave rise to a wave of distrust among banks and a consequent liquidity drought. 

\begin{figure}
\includegraphics[width=0.4\textwidth , trim=0cm 0.0cm 0cm 0cm, clip=true]{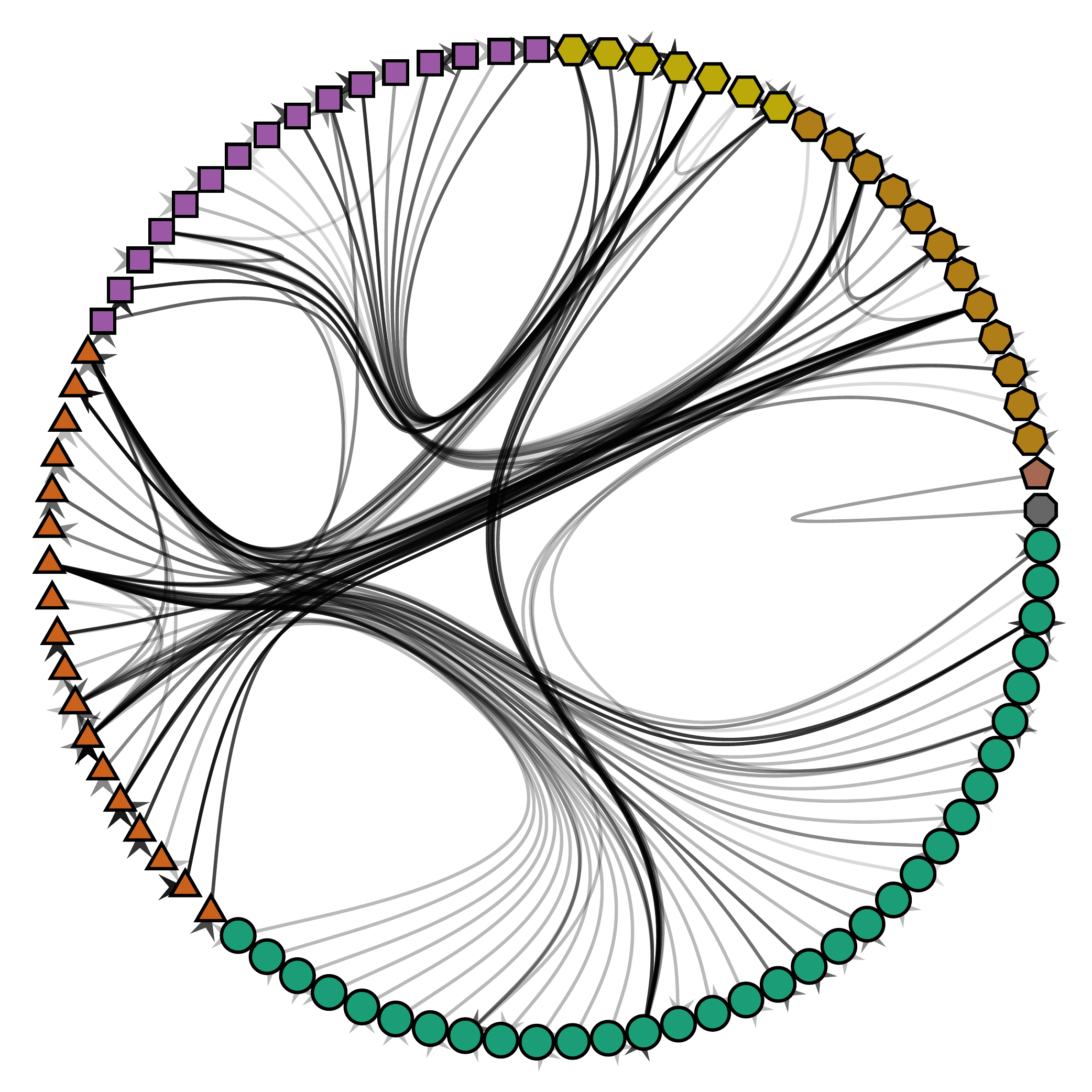}
\includegraphics[width=0.4\textwidth , trim=0cm 0cm 0cm 0.0cm, clip=true]{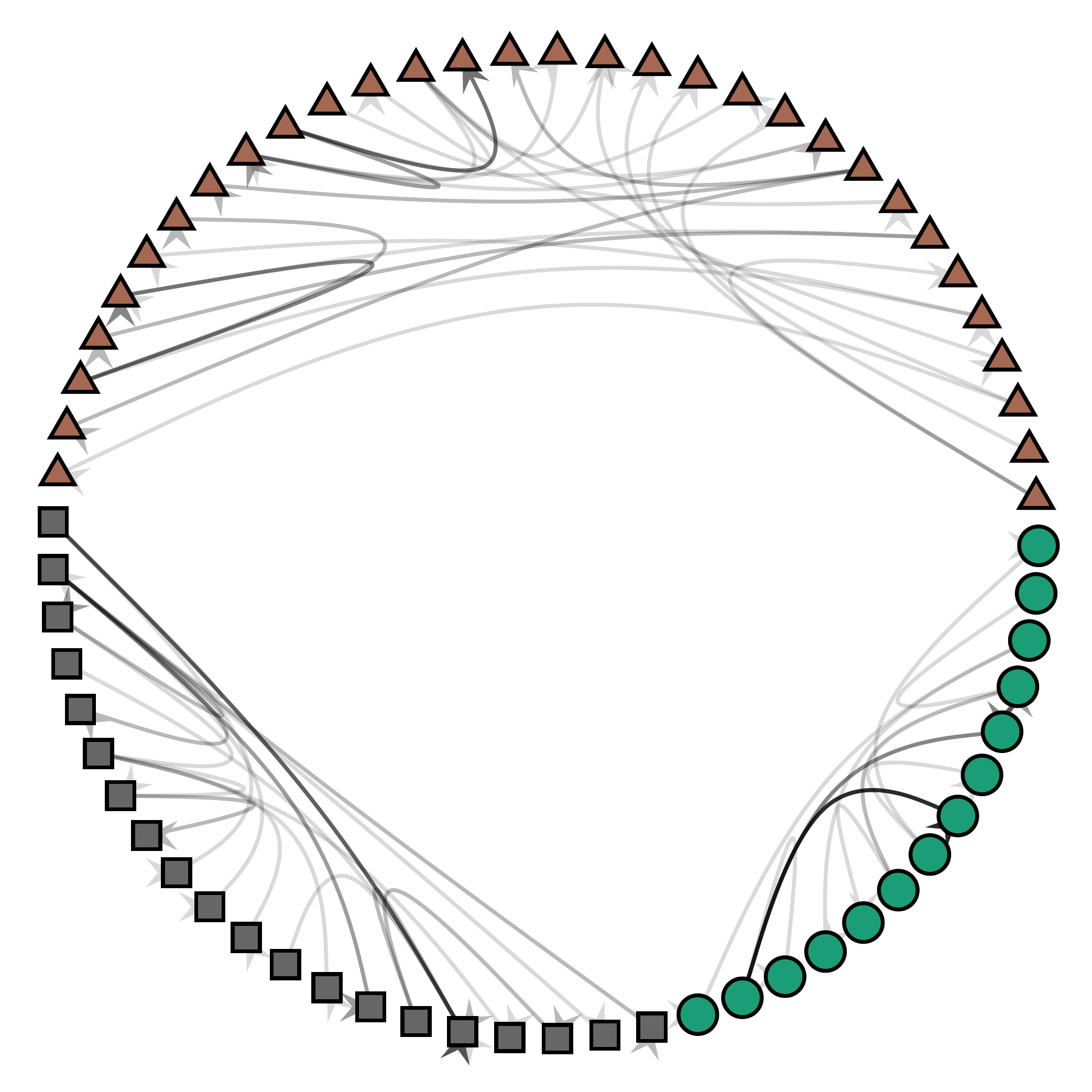} 

\caption{(Color online) The activity on the interbank market in the two weeks leading up to August 17, 1998 (left panel) and the two consecutive weeks (right panel). A bank is represented as node and each link represents an issued loan. Nodes are colored per block as defined in Ref.~\cite{peixoto2014hierarchical}.} 
\label{fig:network_visuals}
\end{figure}

Fig.~\ref{fig:network_visuals} illustrates the impact of Aug 1998 crisis on the Russian interbank network. Nodes are banks, and directed edges represent issued loans. The first panel shows the activity in the market in the two weeks leading up to the seizure, whereas the second panel covers the two weeks after the collapse. Evidently, we see a decrease in the number of active banks, from 87 to 65, and in the number of loans granted, from 507 to 96. When considering the structure in this unweighted multidirected network we can find a clear distinction between the ``normal'' and ``crisis'' periods.  The technique of Ref.~\cite{peixoto2014hierarchical} can be used to uncover groups of nodes, called blocks, which fulfil similar roles in the network 
\footnote{The technique of Ref.~\cite{peixoto2014hierarchical} makes use of a hierarchical version of stochastic block modeling and offers an alternative to the usual modularity optimization algorithms.}.
In the two weeks leading up to the crisis we get an interbank market with $5$ blocks (and two isolated banks). After the crisis the number of identified blocks goes down to $3$. Notably, the identified blocks interact with each other before the crisis but not after. So the crisis disintegrates the network and reconfigures banks' role in it.

We can also see evidence of different development phases by studying the time evolution of two basic network measures: the average local clustering coefficient, which is a potential indicator of systemic risk~\cite{tabak2014directed}, and the average shortest path length. For an undirected network, the local clustering coefficient $0 \le c_{u} \le 1$ of a node $u$  is the number of the edges between the nodes of the neighborhood of $u$ divided by the maximum amount that could possibly exist between them. It can be conveniently defined as in \cite{watts1998collective}
\begin{equation}
c_{u} = \frac{2 T(u)}{deg(u) \left( deg(u)-1 \right) },
\end{equation}
where $T(u)$ is the number of triangles (a subgraph with three nodes and three edges) attached to $u$ and $deg(u)$ is the degree of $u$.
The average of this local clustering coefficient is
\begin{equation}
C = \frac{1}{n}\sum_{u \in G} c_u,
\end{equation}
where $n$ is the number of nodes in the network $G$ \cite{saramaki2007generalizations}.
The average shortest path length $D$ is defined as
\begin{equation}
D =\sum_{u,v \in V} \frac{d(u, v)}{n(n-1)}
\end{equation}
where $V$ is the set of nodes in the network $G$, and $d(u, v)$ is the length of the shortest path from node $u$ to $v$ \cite{NewmanReview}. In a disconnected graph $D$ can only be computed for the largest connected component.

\begin{figure}
\begin{center}
\includegraphics[width=0.5\columnwidth, trim=0cm 0cm 0cm 0cm, clip=true]{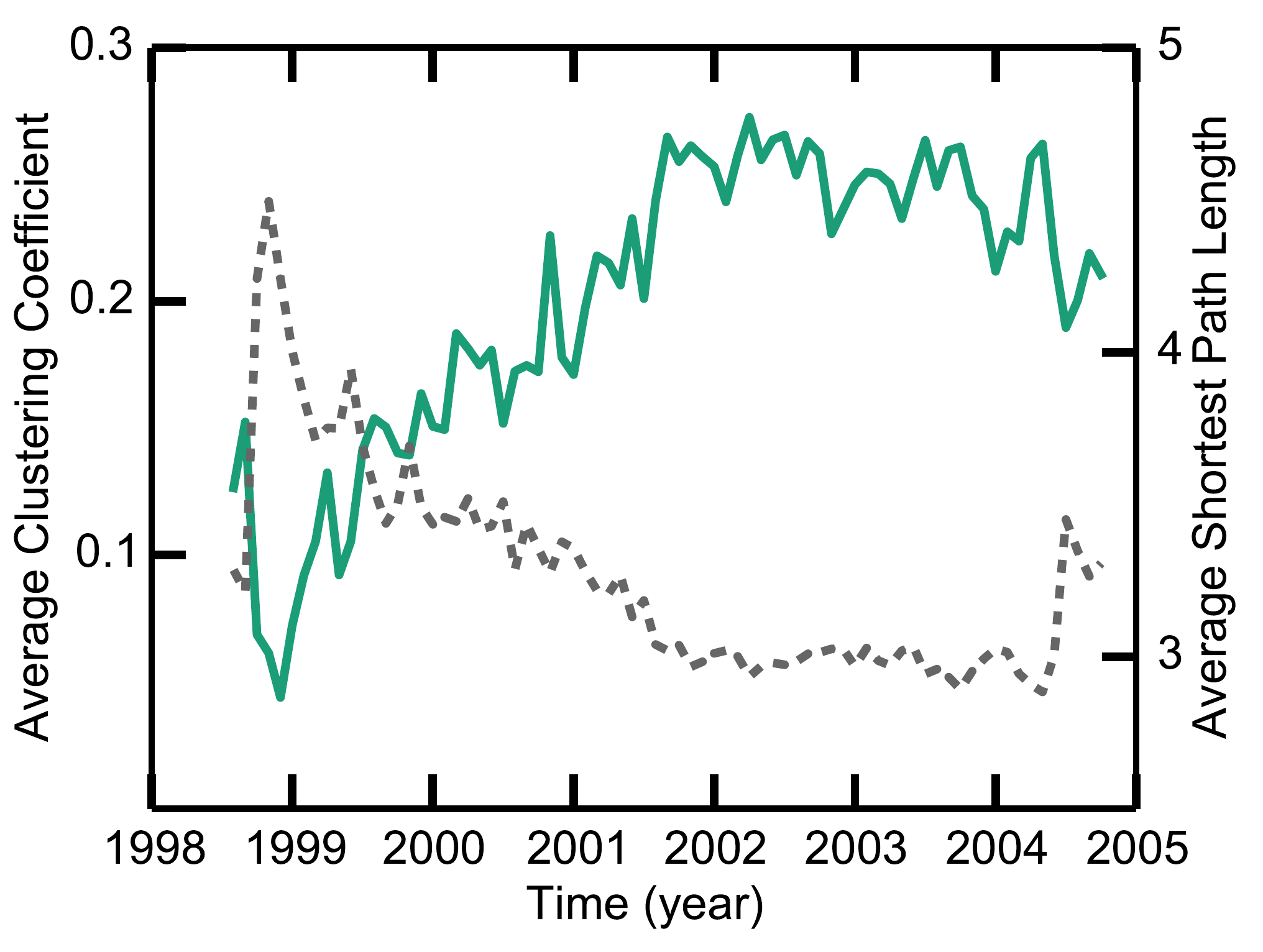}
\caption{(Color online) The time dependence of the average clustering coefficient (solid line) and the average shortest path length (dashed line) calculated for the most connected component of the undirected version of the Russian interbank network between 1998 and 2005. Data are aggregated over a month.}
\label{fig:ClusteringAndASPL}
\end{center}
\end{figure}

Fig.~\ref{fig:ClusteringAndASPL} shows the time evolution of $C$ and $D$ computed per month for the largest connected component of the undirected interbank network. First, note that the local clustering coefficient averaged over all nodes and time periods ($C=0.198$) is nearly identical to the one reported by \cite{cont2013network} for the Brazilian interbank market ($C=0.2$) but is much lower than for the German market (directed clustering $C=0.80$) \cite{roukny2014network}. In contrast, the average shortest path length ($D\approx3$) is notably higher compared to German ($D=2.14$) \cite{roukny2014network}, to Austrian ($D=2$) \cite{AustriaAnalysis} and Mexican ($D=1.7$) \cite{martinez2014empirical} interbank networks. In the case of the Mexican and Austrian ones, this difference is likely driven by the fact that the Russian network is sizably larger. 

The $C$ and $D$ tend to move in the opposite direction: three months after the first crisis hits $C$ drops while $D$ spikes; during the growing phase of the network (1999-2002) $C$ grows while $D$ falls; during the mature phase (2002-2004) both measures stabilize at $C\approx 0.22$ and $D\approx 3$; finally, during the 2004 crisis $C$ drops while $D$ spikes again. The average local clustering coefficient $C$, however, tends to have bigger fluctuations from period to period. In contrast, the time series of the average shortest path length is very smooth, and the only two obvious spikes occur around the two crises. Clearly, those crises disrupted the overall network structure: the liquidity drought, which is equivalent to the pruning of links, caused a significant increase in the average shortest path length. Even the decrease in the number of nodes, and hence shrinking of the network, could not offset this effect. The 1998 spike in $D$ is particularly remarkable given we only consider the largest connected component, that is, about one third of the nodes (see Fig.~\ref{fig:network_visuals}).

\section{Methodology}\label{Sec:methodology}
For each network measure we test whether the power law or an alternate fat-tailed distribution fits the data best. First, we fit distributions to the tail, then to the entire data range. Along the way, we apply a criterion which allows us to discriminate between the tail and the bulk parts of the data.

To study the tail we adopt the methodology of Ref.~\cite{ClausetPL}. First, we fit a power law (PL) to the data to determine the starting point of the tail part (the so-called cut-off ${x}_{min}$ of the scaling range). Then we fit each of the theoretical distributions to the tail using maximum likelihood (ML). Finally, we use a relative goodness-of-fit test to select the distribution that fits the data best. We consider the same selection of theoretical distributions as in Ref.~\cite{ClausetPL}. Their functional forms are listed in Table~\ref{tab:DefinitionOfDistributions} and involve one or two free parameters.

\begin{table}
\caption{Definition of the normalized distributions used in this
  work. The constant $C$ is defined by the normalization condition
  $\int_{x_{min}}^{\infty}  f(x) dx = 1$. }
\begin{tabular}{ l  c }
\hline \hline
  Distribution 			& $f(x)$ \\
  \hline
  Power law (PL)			& $Cx^{-\alpha}$  \\
  Truncated power law (TPL) 	& $Cx^{-\alpha} e^{-\lambda x}$  \\
  Exponential 	(Exp)		& $Ce^{-\lambda x}$ \\
  Stretched exponential (SExp) & $Cx^{\beta - 1} e^{-(\lambda x)^{\beta}}$ \\
  Log-normal (LN)	& $ \frac{C} {x} \exp \left[ - \frac{ \left( \ln x - \mu \right)^{2} }{2 \sigma^{2}} \right] $ \\
\hline \hline
\end{tabular}
\label{tab:DefinitionOfDistributions}
\end{table}

For completeness and to introduce the notation we briefly sketch the methodology of Ref.~\cite{ClausetPL}. Consider a given ordered data set $\{ x_{j},j=1,\ldots,N\}$. Every entry $x_{j}$ is a potential $x_{min}$ and for each of those we compute the ML estimate of the power-law exponent $\alpha$
\begin{equation}
\hat{\alpha} \left( x_j = x_{min} \right) = 1 + (N-j+1) \left[ \sum_{i=j}^{N} 
\ln \frac{ x_{i} }{ x_{min} }  \right] \; .
\end{equation}

We then use the Kolmogorov-Smirnov test to select the optimum $\hat{x}_{min}$. It is defined as the cut-off which minimizes the quantity   
\begin{equation}
Z = \max_{x \geq x_{min}} \mid S(x) - P(x) \mid.
\label{eq:KScriterion}
\end{equation}
Here, $S(x)$ is the cumulative distribution function (CDF) of the observed values for $x_j \geq x_{min}$, and $P(x)=\frac{C}{- \alpha +1} \left( x^{-\alpha +1}-x_{min}^{-\alpha+1} \right)$ is the CDF of the power-law fit to the tail part of the data. 

In the next step, for given $x_{min}$, we perform ML fits to the data with the other four candidate distributions of Table~\ref{tab:DefinitionOfDistributions}. As argued by Ref. ~\cite{ClausetPL}, it is more useful to know which distribution is the best possible fit candidate, rather than the goodness of fit for each distribution individually. To compare the relative goodness of the different fits, we compute the likelihood ratios $R$ for the pairs of probability density functions (PDFs) $p_{1}(x_{i})$ and $p_{2}(x_{i})$
\begin{equation}
R \left( p_{1}, p_{2} \right) = \frac{L_{1}}{L_{2}} = \prod_{i=j}^{N} \frac{p_{1}(x_{i})}{p_{2}(x_{i})} \; .
\end{equation}
The corresponding normalized loglikelihood ratios $\mathcal{R}\left( p_{1}, p_{2} \right)$ read
\begin{eqnarray} 
\mathcal{R} \left( p_{1}, p_{2} \right) &=& \frac{1}{\sigma _{12} \sqrt{N-j+1}} \sum_{i=j}^{N} \left[ \ln \frac {p_{1}(x_{i})} {p_{2}(x_{i})} \right] \nonumber \\
&=&  \frac{1}{\sigma_{12} \sqrt{N-j+1}} \sum_{i=j}^{N} \left[ l_{i}^{(1)} - l_{i}^{(2)} \right] ,
\label{eq:ratiologlike}
\end{eqnarray}
where $l_{i}^{(k)} = \ln p_{k}(x_{i})$ and $\sigma _{12}$ is defined as
\begin{equation}
\sigma_{12}^{2} = \frac{1}{N-j+1} \sum_{i=j}^{N}  \left[ \left( l_{i}^{(1)} - l_{i}^{(2)} \right)  -  \left( \bar{l}^{(1)} - \bar{l}^{(2)}  \right) \right]^{2}.
\end{equation}
The $\mathcal{R} \left( p_{1}, p_{2} \right)$ is positive (negative) if the data is more likely in the $p_{1}$ ($p_{2}$) distribution. 

In order to guarantee that the value of $\mathcal{R}$ is not merely a product of fluctuations and that the true expectation value of $\mathcal{R}$ is zero, we compute the probability that the measured normalized log likelihood ratio has a magnitude as large or larger than the observed value $\mathcal{|R|}$. This so-called $p$-value is defined as
\begin{equation}
p \left( \mathcal{R}  \right)= \frac{1}{\sqrt{2\pi}} \left[ \int_{-\infty}^{-|\mathcal{R}|} e^{-\frac{t^{2}}{2}} dt + \int_{|\mathcal{R}|}^{\infty} e^{-\frac{t^{2}}{2}} dt \right].
\end{equation}
The distribution $p_{k}$ with the highest value of 
\begin{equation}
g \left( p_{k} \right) = \sum _{l \ne k} \mathcal{R} \left( p_{k}, p_{l}
\right)  \; ,
\label{eq:sumofRs}
\end{equation}
is considered as the most suitable distribution among the different  candidates $ \left\{ p_{i} \right\}$. Bootstrapping and the Kolmogorov-Smirnov test are
alternate methods to compute the $p$ values. Both methods, however, are
subject to some pitfalls as outlined in Ref.~\cite{Alstott2014powerlaw}. By using the PL as a benchmark and letting it decide on the value of $x_{min}$, we are confident that our methodology produces the best possible PL fit to the data. If it turns out that  an alternate distribution provides a significantly better fit to the tail part of the data than the PL, strong evidence emerges that this theoretical distribution better accounts for reality than a power law.

We stress that the above-sketched methodology of Ref.~\cite{ClausetPL} can not only be applied to the analysis of the tail part of the data but also to the full range of the data by setting $x_{min}=x_1$. When considering the full range of data it is natural to select a set of theoretical distributions that encompass a Gaussian-like regime (thermal or bulk part) supplemented with a fat tail (superthermal part). Examples of such distributions include the stretched exponential and the log-normal.

\begin{figure}
\includegraphics[width=0.45\columnwidth, trim=0cm 0cm 0cm 0cm, clip=true]{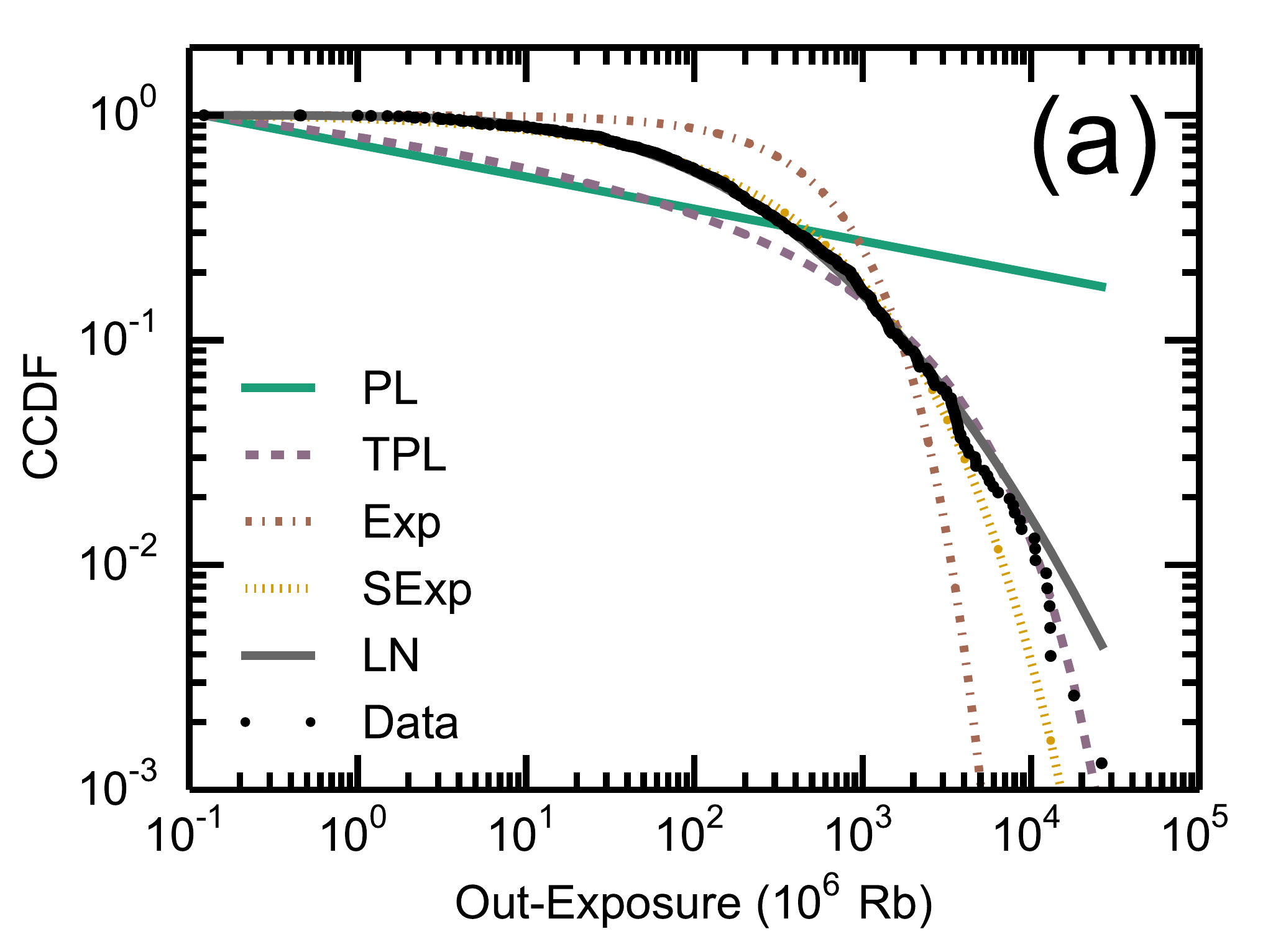} 
\includegraphics[width=0.45\columnwidth, trim=0cm 0cm 0cm 0cm, clip=true]{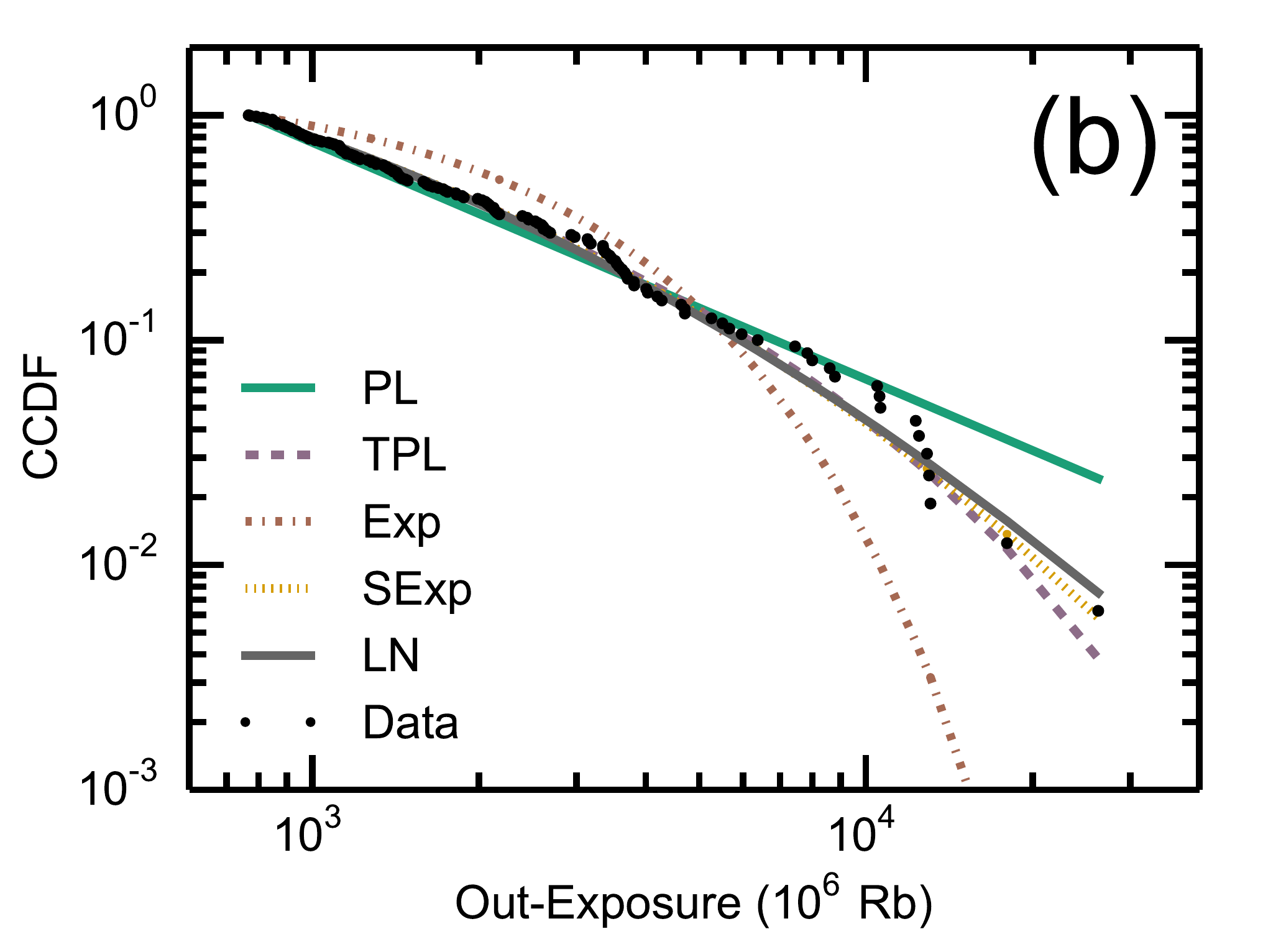} 
\caption{(Color online) The complementary cumulative distribution functiom (CCDF) of the entire (a) and tail part (b) of the  monthly-aggregated out-exposure
  distribution for January 2002. Also shown are the ML fits to the CCDFs for the
  five distributons of Table~\ref{tab:DefinitionOfDistributions}. 
}
\label{fig:Fits}
\end{figure}

\begin{figure*}
\includegraphics[width=1.\textwidth , trim=0cm 2.115cm 0cm 0cm, clip=true]{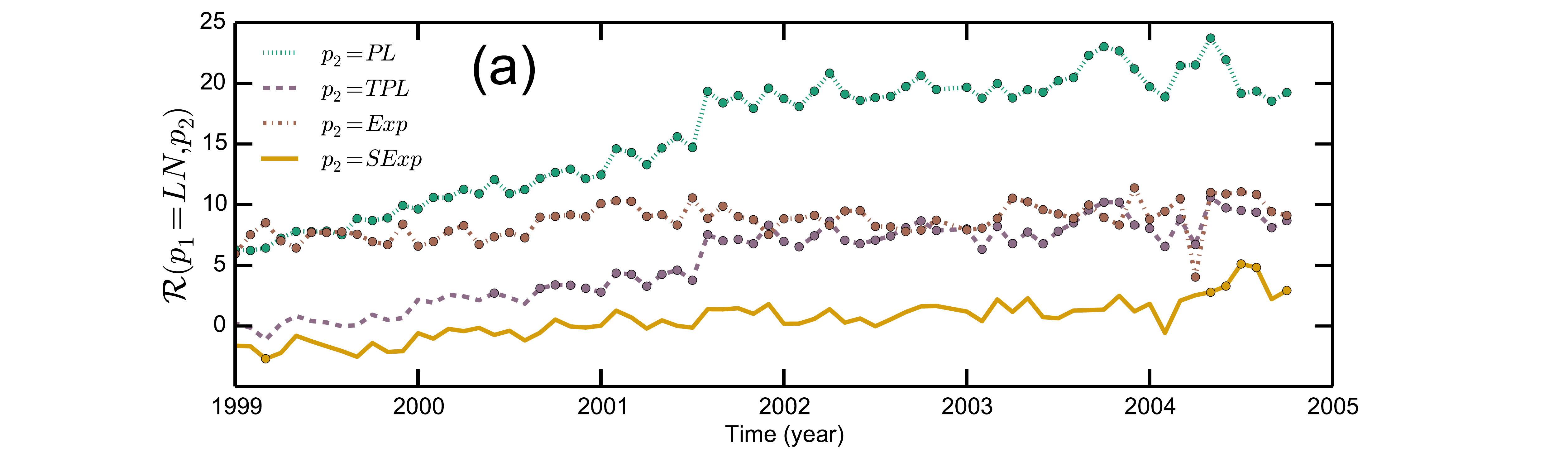} \\
\includegraphics[width=1.\textwidth , trim=0cm 0cm 0cm 0cm, clip=true]{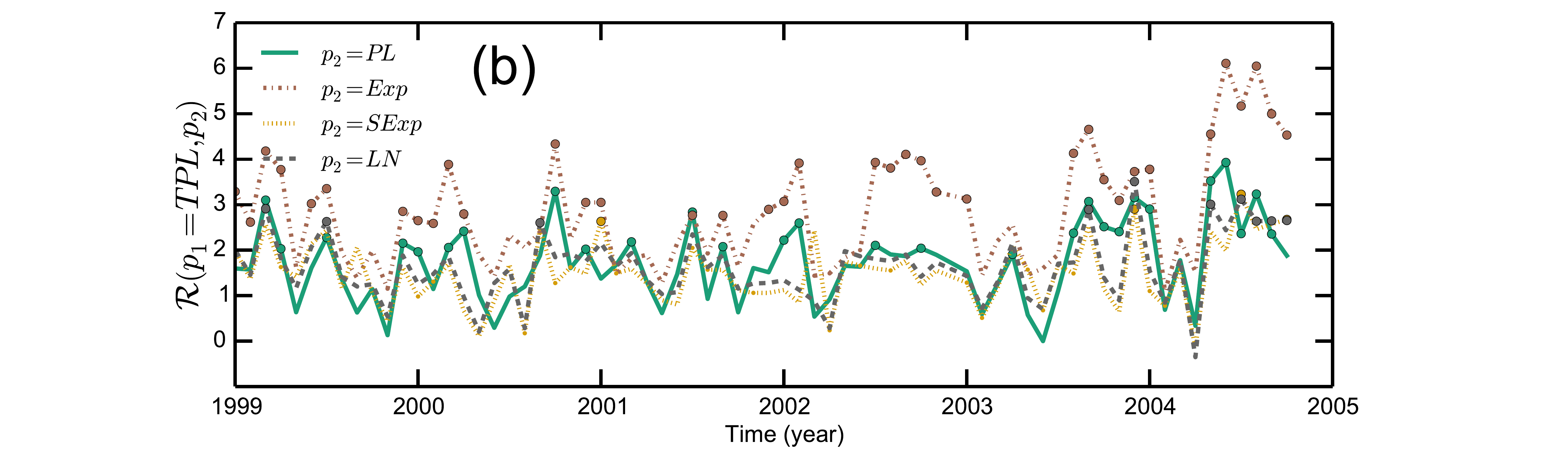} 
\caption{(Color online) (a) The time series of the normalized loglikelihood ratios
$\mathcal{R} \left( p_{1}=\text{LN}, p_{2} \right)$ of Eq.~(\ref{eq:ratiologlike}) for the
  different fits to the total distribution of the monthly-aggregated
  out-exposure. (b) The time series of $\mathcal{R} \left(
  p_{1}=\text{TPL}, p_{2} \right)$ for the fit to the tail of the 
  monthly-aggregated out-exposure distribution. }
\label{fig:LR}
\end{figure*}

\begin{figure*}
{\includegraphics[width=1.\textwidth , trim=0cm 2.115cm 0cm 0cm, clip=true]{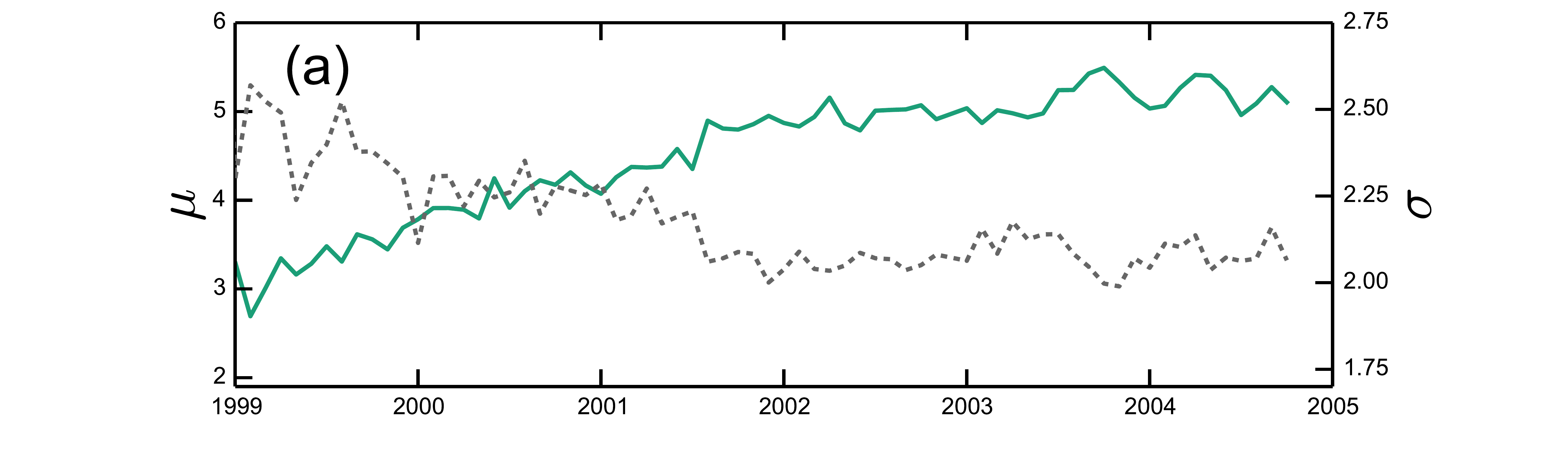}} \\
{\includegraphics[width=1.\textwidth , trim=0cm 0cm 0cm 0cm, clip=true]{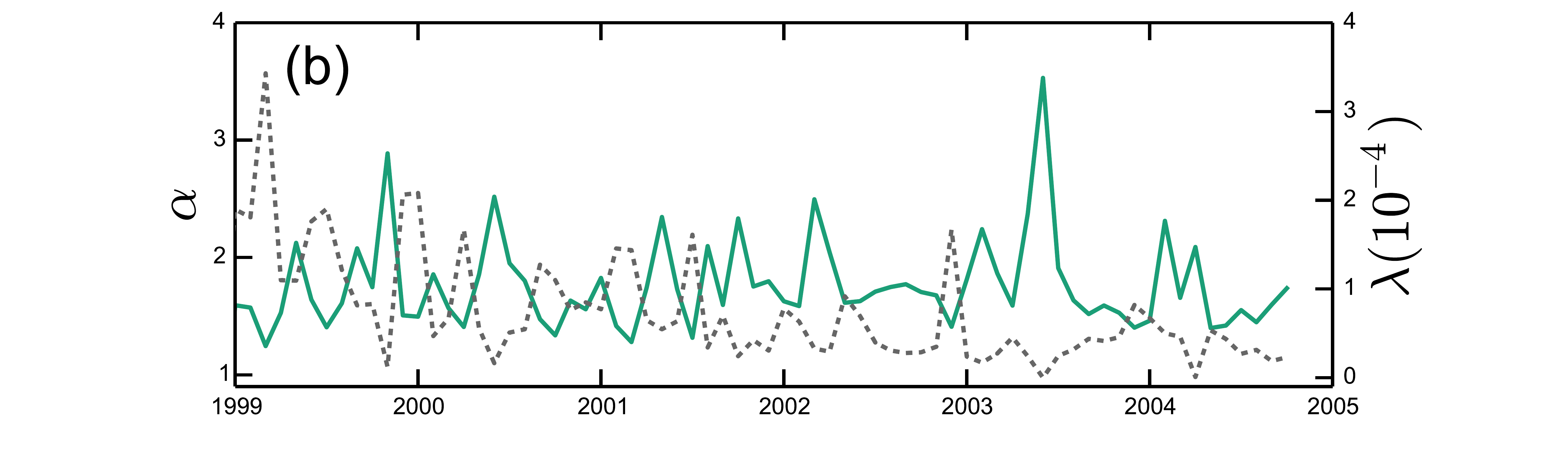}} 
\caption{(Color online) The time evolution of the parameters for the best fit
  to the total (a) and tail (b) part of the out-exposure distributions. The total data range is fitted with a log-normal distribution
  with parameters $\mu$ (solid line) and $\sigma$ (dashed line). The
  tail part is fitted with a truncated power-law
  distribution with parameters $\alpha$ (solid line) and $\lambda$ (dashed line).}
\label{fig:parameters}
\end{figure*}

\begin{table*}
\caption{Summary of the results of the fits to the tails of the monthly-aggregated
  data of all the network features considered in this work.  We list the
  average fraction of the events in the tail and the distribution
  which provides the best fit, together with the percentage of months
  for which it comes out ``best'' according to the criterion of
  Eq.~(\ref{eq:sumofRs}). Columns 4 and 5 provide the average values
  of the parameters of the best-fit distributions. The last column lists the distributions
  which are not significantly (1 percent level) worse for more than
  half of the 70 months (January $1999$ - October $2004$) considered, together with the percentage of
  months for which this is the situation. The measures with an extra label "(B)" refer to an analysis of a sample obtained after bootstrapping the real data.}
\begin{tabular}{ l  c  c  c  c  c }
\hline \hline
	Measure 					& 	Tail (\%) & Best fit (\%) & $\alpha$ & $\lambda$ &	Alternate fits (\%) \\
	\hline
	Asset Size	 				&  	$23 \pm 9$ 	& TPL (77) & $1.80 \pm 0.06$ & $(4.47 \pm 9.56) \times 10^{-7}$ & PL (100), SExp (100), LN (97) \\
	Capital Size  				&	$31 \pm 18$ & TPL (58) & $1.91 \pm 0.14$ & $(1.25 \pm 1.71) \times 10^{-5}$ & PL (100), SExp (100), LN (89) \\
    Leverage 					&	$24 \pm 9$	& PL (57)  & $3.67 \pm 0.76$ & None 							& TPL (100), Exp (86), SExp (100), LN(100) \\

Leverage (B) 					&	$28 \pm 10$ & TPL (57)  & $3.77\pm0.10$ & $(1.81\pm2.77) \times 10^{-6} $							& PL (100), Exp (93), SExp (100), LN(100) \\

	Loan Size 					&  	$5 \pm 9$ 	& TPL (75) & $2.49 \pm 0.47$ & $(3.22 \pm 3.54) \times 10^{-4}$ & PL (91), SExp (87), LN (71) \\
	Counterparty-Exposure		& 	$12 \pm 8$ & TPL (80) & $2.17 \pm 0.51$ & $(2.53 \pm 2.44) \times 10^{-4}$ & PL (53), SExp (81), LN (71) \\
		
	Counterparty-Exposure (B)		& 	$16\pm9$ & TPL (86) & $2.02\pm0.36$ & $(3.13\pm2.33) \times 10^{-4}$ & PL (91), SExp (100), LN (100) \\
		
Undirected Degree 			& 	$21 \pm 16$ & TPL (100) & $1.80 \pm 0.36$ & $(1.83 \pm 0.83) \times 10^{-2}$ & PL (70), Exp (68), SExp (97),  LN (97) \\
	Directed In-Degree			& 	$25 \pm 20$ & TPL (94) & $1.72 \pm 0.43$ & $(2.65 \pm 2.74) \times 10^{-2}$ & PL (71), Exp (68), SExp (91), LN (95) \\
	Directed Out-Degree 		& 	$13 \pm 8$ & TPL (89) & $2.07 \pm 0.66$ & $(2.93 \pm 1.92) \times 10^{-2}$ & PL (85), Exp (98), SExp (100), LN (100) \\
    Directed Out-Degree (B)		& 	$28\pm18$ & TPL (77) & $1.94\pm0.71$ & $(4.13\pm3.99) \times 10^{-2}$ & PL (99), Exp (100), SExp (99), LN (99) \\
	Multidirected In-Degree 	& 	$23 \pm 24$ & TPL (97) & $1.83 \pm 0.48$ & $(5.05 \pm 3.93) \times 10^{-3}$ & PL (74), Exp (77), SExp (99), LN (100) \\
	Multidirected Out-Degree	& 	$23 \pm 17$ & TPL (93) & $1.74 \pm 0.50$ & $(7.80 \pm 5.83) \times 10^{-3}$ & PL (63), Exp (88), SExp (93), LN (91) \\
	
    In-Exposure		 			& 	$30 \pm 17$ & TPL (97) & $1.49 \pm 0.43$ & $(7.68 \pm 5.11) \times 10^{-5}$ & SExp (57) \\
	Out-Exposure 				& 	$20 \pm 11$ & TPL (94) & $1.76 \pm 0.39$ & $(7.49 \pm 6.23) \times 10^{-5}$ & PL (59), SExp (94), LN (83) \\

\hline \hline
\end{tabular}
\label{tab:ResultsTailsv2metJ1}
\end{table*}

\section{Illustration of Methodology for Out-exposure distribution}\label{Sec:illustration}
As a prototypical example of the methodology sketched in Sec.~\ref{Sec:methodology}, we explore the out-exposure distribution for the monthly aggregated data.  The data for the out-exposure distribution of January 2002 shown in Fig.~\ref{fig:Fits} are exemplary for all considered 70 months (January 1999 - October 2004). The considered values for the out exposure extend over more than five orders of magnitude and the distribution can be labelled as heavy-tailed~\footnote{A distribution is defined as heavy-tailed if it is not exponentially bounded \cite{Alstott2014powerlaw}.}.

Using the methodology outlined in Sec.~\ref{Sec:methodology} we identify the tail part of the data. From Fig.~\ref{fig:Fits}(b) it is clear that fits to the tail part of the considered data with the two-parameter log-normal (LN), stretched exponential (SExp) and truncated power law (TPL) distributions outperform those with the one-parameter exponential (Exp) and power law (PL) distributions. The parameter $\lambda$ in the TPL accounts for the finite-size effects near the upper edge of the out-exposure distribution.  For each of the 70 months in our sample we compute the ratios $\mathcal{R}\left(p_{1}, p_{2} \right)$ of Eq.~(\ref{eq:ratiologlike}) for all pair combinations out of the list of five distributions of Table~\ref{tab:DefinitionOfDistributions}. In Fig.~\ref{fig:LR} we display the time series of the $\mathcal{R} \left( \text{TPL}, p_{2} \right)$ for $p_{2}=$PL, Exp, SExp, and LN.  We observe that $\mathcal{R} \left( \text{TPL}, p_{2} \right)$ is mostly positive, which indicates that the TPL offers the best overall description of the tail of the out-exposure data. In order to test the significance of this observation, we evaluate the $p$ values. If the normalized ratio for a pair of distributions in a given month is significant at a one percent level, the data points in Fig.~\ref{fig:LR} are dotted. The figure indicates that the two-parameter TPL is a significantly better fit to the tail of the monthly-aggregated out-exposure distribution than a power law and an exponential. The TPL, however, does not provide a significantly better fit than the LN or SExp for most of the months. 

Fig.~\ref{fig:parameters}(b) shows the time evolution of the TPL parameters. Whereas $\alpha$ fluctuates around the same average value, $\lambda$ falls over time. Hence, as time proceeds, the exponential cutoff to the power law shifts to larger values of the out-exposure.

To weigh the relative performance of the different distributions we compute their monthly $g \left( p_{i} \right)$ scores (see Eq.~(\ref{eq:sumofRs})). We dub $p_{i}$ as the ``best'' overall fit candidate when its $g \left( p_{i} \right)$ score is highest for the largest fraction of the 70 months in our sample.  As can be seen in Table~\ref{tab:ResultsTailsv2metJ1}, the truncated power law is the best-fit candidate in $94$\% of the considered months.  Nevertheless, $\mathcal{R} \left( \text{TPL}, \text{SExp} \right)$ is not significantly different from zero in $94$\% of the considered months as is the case with $\mathcal{R} \left( \text{TPL}, \text{PL} \right)$ in $59$\% and with  $\mathcal{R} \left( \text{TPL}, \text{LN} \right)$ in $83$\% of the months. So we list the SExp as well as PL and LN as alternate best-fit candidates. In general, if a distribution is not significantly worse than the best-fit candidate in more than half of the months, it is mentioned in the last column of Table~\ref{tab:ResultsTailsv2metJ1}. We conclude that although the tail is described best by a TPL, the fit is not significantly better than the power-law, stretched-exponential and log-normal fits.

We now turn our attention to the distributions covering all out-exposure data points, from small to large. Again we consider the five PDFs of Table~\ref{tab:DefinitionOfDistributions}. From the ML fits, for example for January 2002 displayed in Fig.~\ref{fig:Fits}(a), we immediately notice that the exponential and power law do not fit the entire range of data well. This is a clear indication of the fact that the distribution of the out-exposure has both a thermal (Gaussian-like) and a superthermal (fat tail) part. Fig.~\ref{fig:LR} shows the normalized loglikelihood ratios $\mathcal{R} \left( p_{1}=\text{LN}, p_{2} \right)$ over time. Using the same criteria as for the tail part, we conclude that the log-normal is the best fit candidate for $80$\% of the 70 months studied.

We also find that the stretched exponential is not significantly worse in $57$\% of the cases, which is reflected in the persistently small values of their likelihood ratios in the top panel of Fig.~\ref{fig:LR}.  During the $2004$ crisis the stretched exponential has a significantly better fit than the log-normal. As the number of interbank loans is subject to a sudden drop in the summer of 2004 (Fig.~\ref{fig:NodeTransactionNumber}), this change in the best-fit candidate illustrates how the network adapts to changing overall conditions. Fig.~\ref{fig:parameters}(a) shows the time evolution of the log-normal fit parameters. The parameter $\mu$ increases during the ``growth'' stage of the network whereas $\sigma$ decreases, and both tend to flatten out in the ``mature'' phase of the network when few nodes or links are added (see Fig.~\ref{fig:NodeTransactionNumber}).
 
\begin{figure}
\begin{center}
\includegraphics[width=0.5\columnwidth, trim=0cm 0cm 0cm 0cm, clip=true]{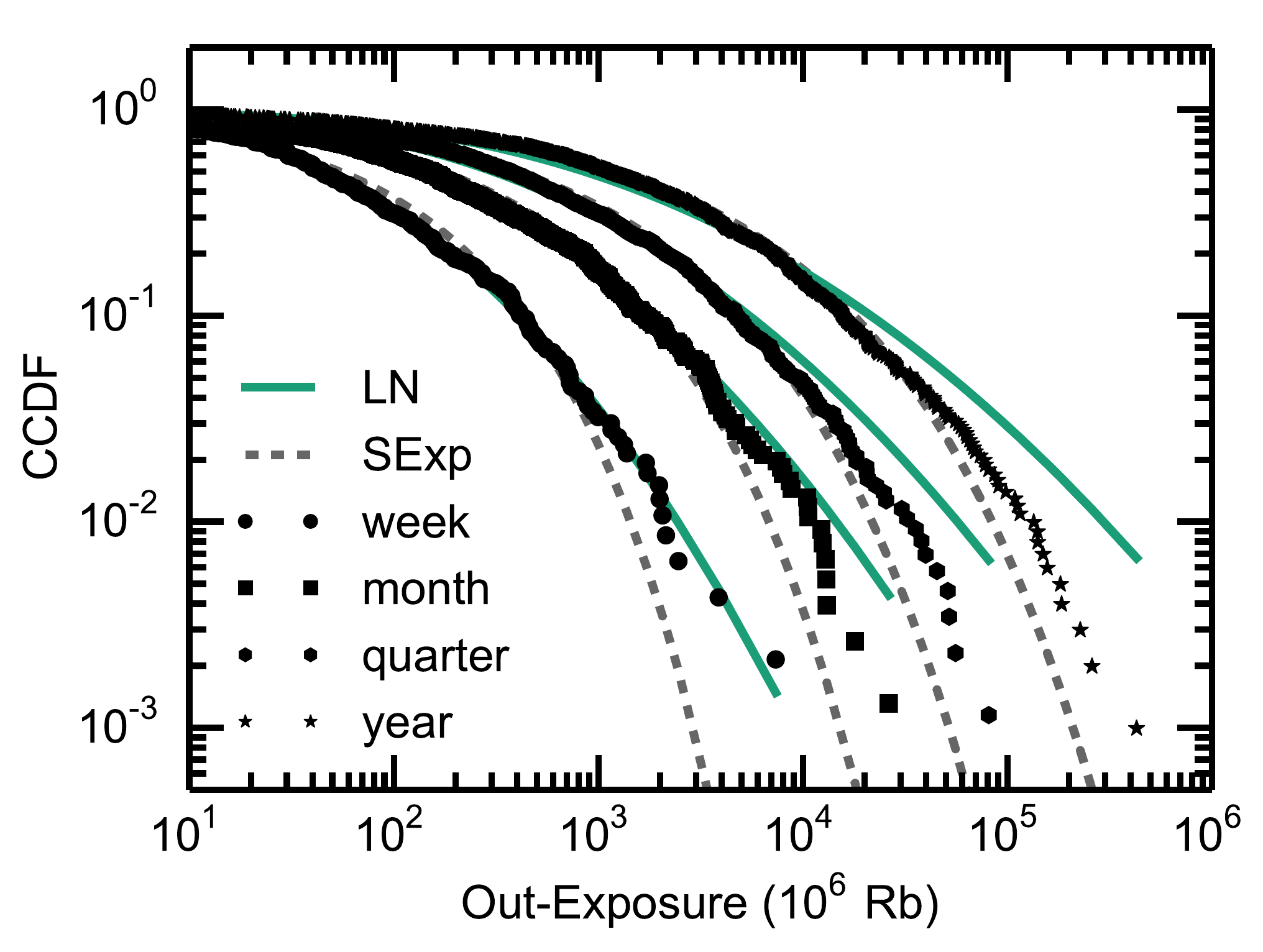}
\caption{(Color online) The CCDF of the out-exposure distribution from a network
  aggregated with data from one week (week 2 of January 2002), from
  one month (January 2002), from one quarter (first quarter of 2002),
  and one year (2002). We also show the respective stretched exponential and
  log-normal fits.}
\label{fig:EOmDifferentAggregates}
\end{center}
\end{figure}

Now, we investigate in how far the size of the time bin widths affects the conclusions with regard to the ``best'' theoretical distributions describing the data. To this end, we study the out-exposure and consider different time intervals to aggregate the data. Beside the networks build with all loans issued in one month, we can do this for all loans issued in a week, quarter or year. Fig. \ref{fig:EOmDifferentAggregates} shows the out-exposure distribution for the second week of January 2002, January 2002, the first quarter of 2002, and the entire year of 2002. We also include the ML fit with the stretched exponential and log-normal for each of the aggregates. If we calculate $g \left( p_{k} \right)$ for every distribution in Table~\ref{tab:DefinitionOfDistributions}, we find that the log-normal is the preferred candidate to fit each dataset, although, again, the stretched exponential is not significantly worse. For the tail part, the truncated power law stays on top too. Because the data has a monthly periodicity, linked to the monthly compliance with regulatory requirements, aggregating the data monthly is the most natural thing to do.

To check whether these observations can be generalized, we repeat the illustrated procedure for weekly, quarterly, and yearly aggregates. Tab.~\ref{tab:test} shows the ``best'' fit candidate for the tail and entire distribution over our data sample. For the tail part, the TPL is preferred over the others for each aggregate. It does so in $85\%$ of the weeks and up to $100\%$ of the years. For the entire distribution, the LN is always preferred.

\begin{figure}
\begin{center}
\includegraphics[width=.45\columnwidth, trim=0cm 0cm 0cm 0cm, clip=true]{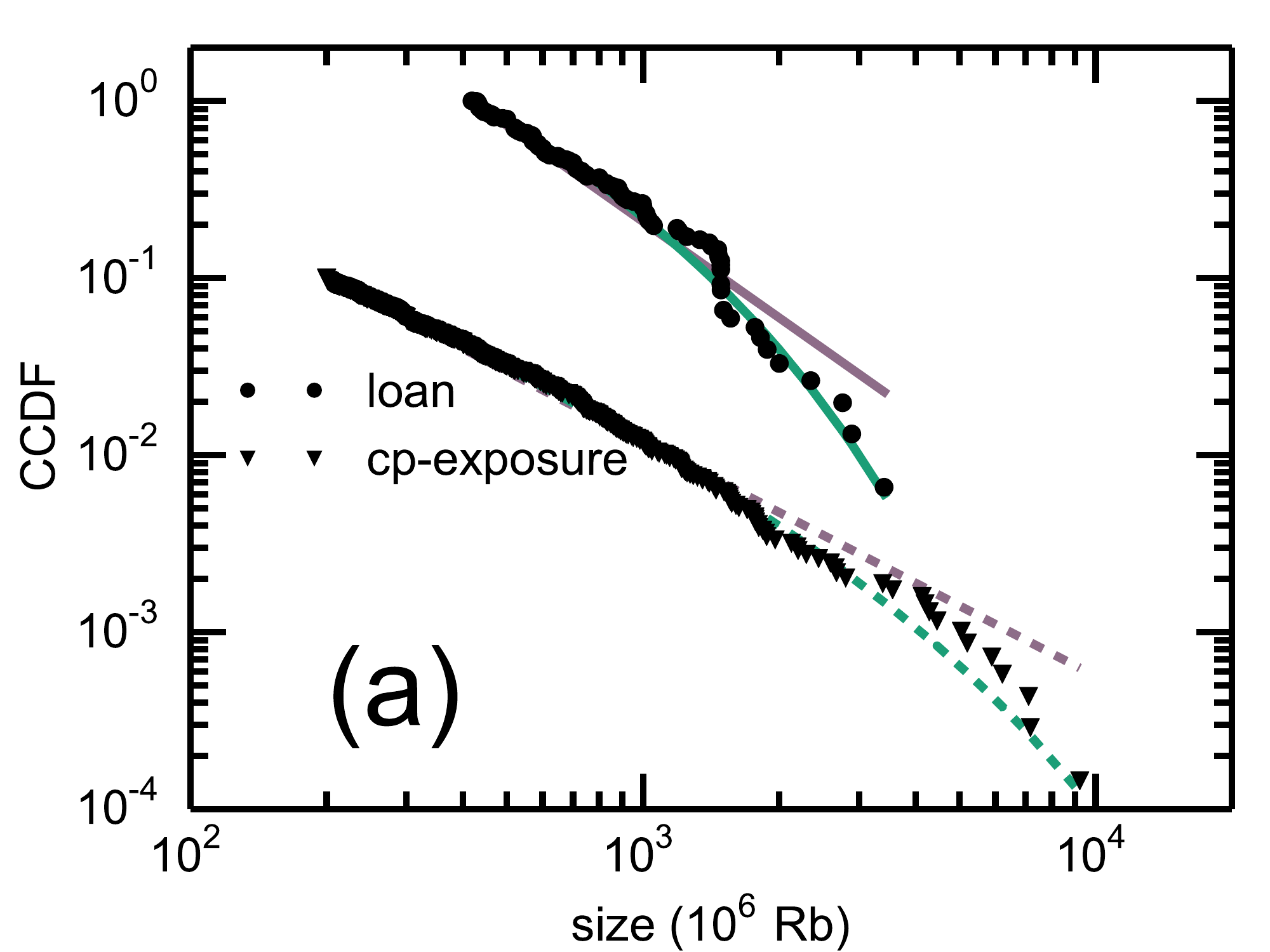} 
\includegraphics[width=.45\columnwidth, trim=0cm 0cm 0cm 0cm, clip=true]{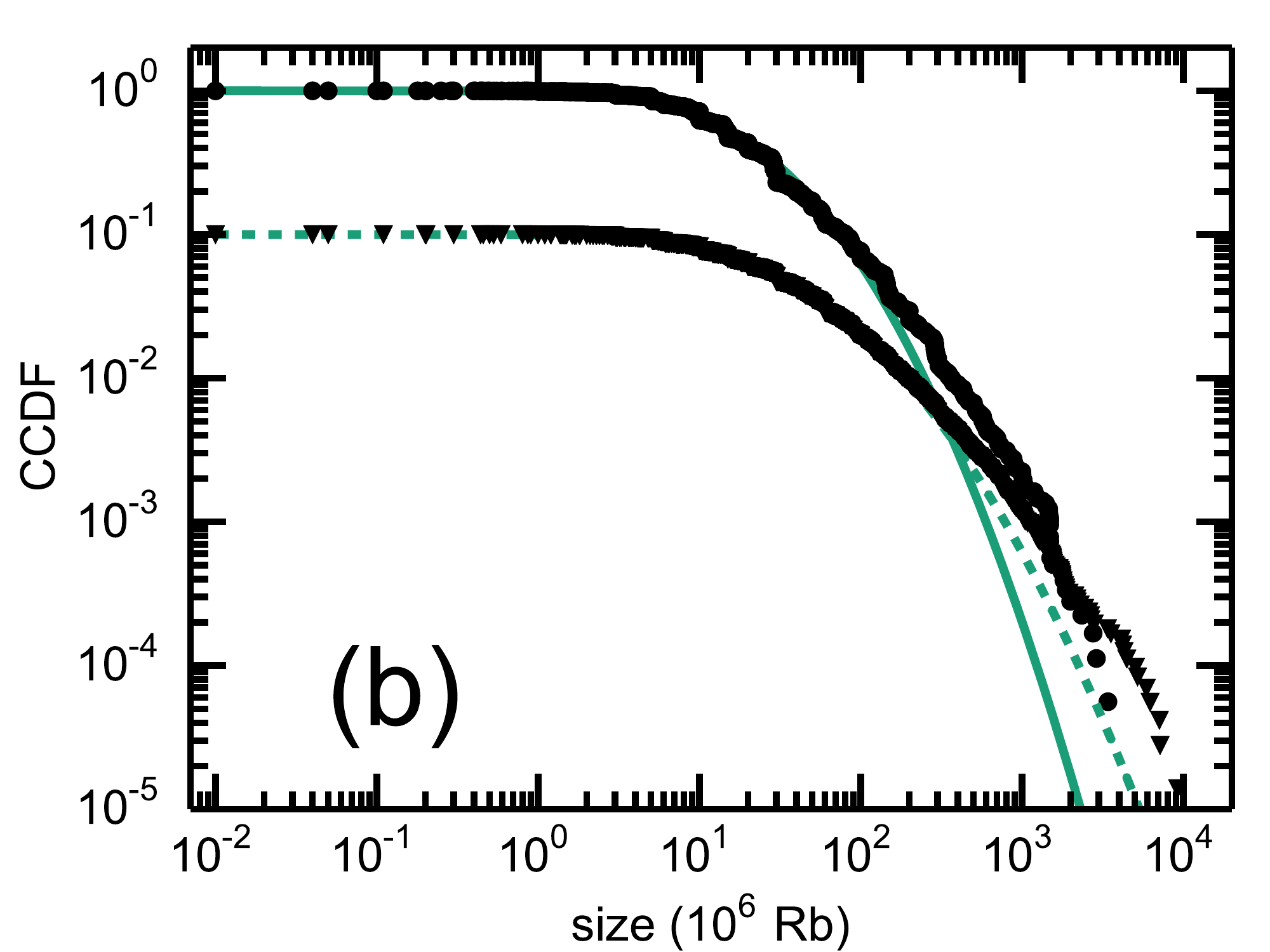}
\caption{(Color online) The CCDF for the loan-size and the counterparty(cp)-exposure for January 2004.  Panel (a) shows the tail part of the data together with a PL (grey) and TPL (green) fit. Panel (b) shows the data over the entire range together with LN fit. For illustrative reasons, the data for the cp-exposure have been scaled by a factor of $0.1$.}
\label{fig:CPandLoan}
\end{center}
\end{figure}

\begin{figure*}
\begin{center}
\includegraphics[width=\textwidth, trim=0cm 0cm 0cm 0cm, clip=true]{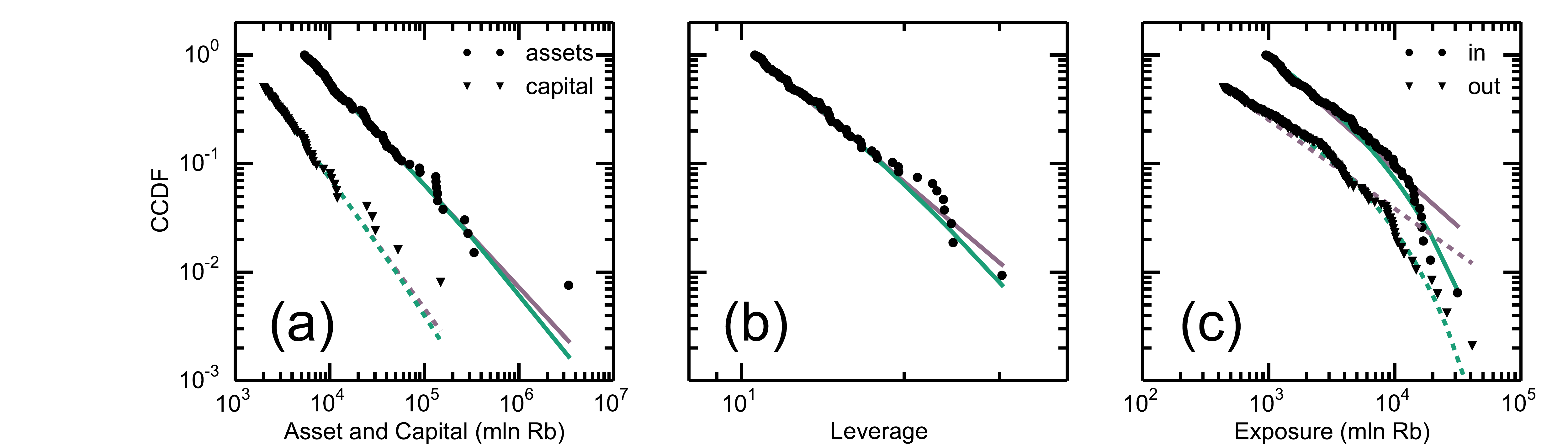}
\includegraphics[width=\textwidth, trim=0cm 0cm 0cm 0cm, clip=true]{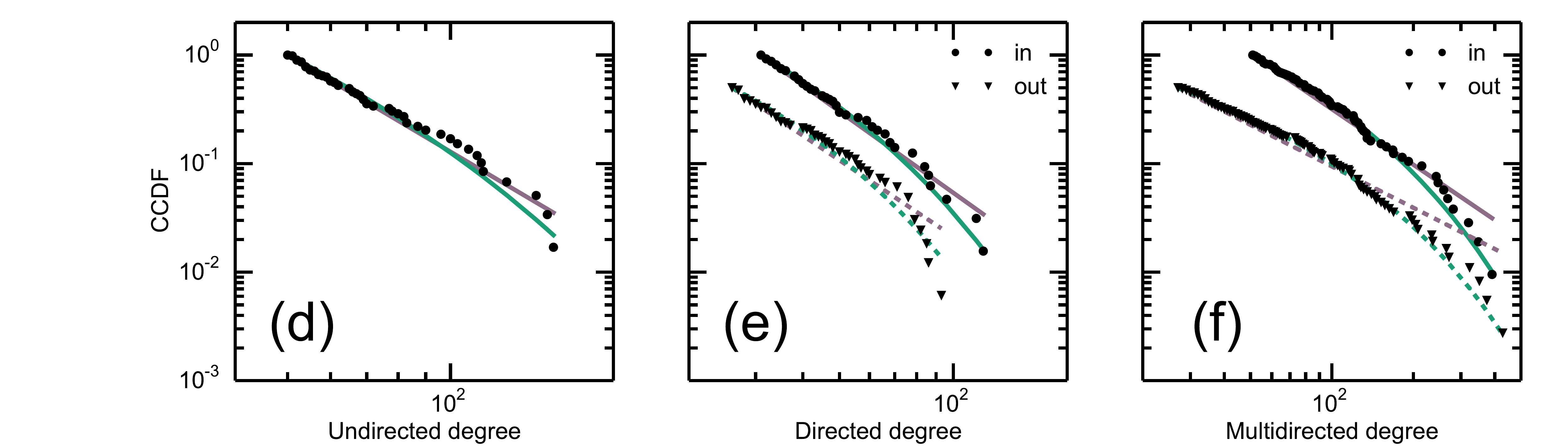}
\includegraphics[width=\textwidth, trim=0cm 0cm 0cm 0cm, clip=true]{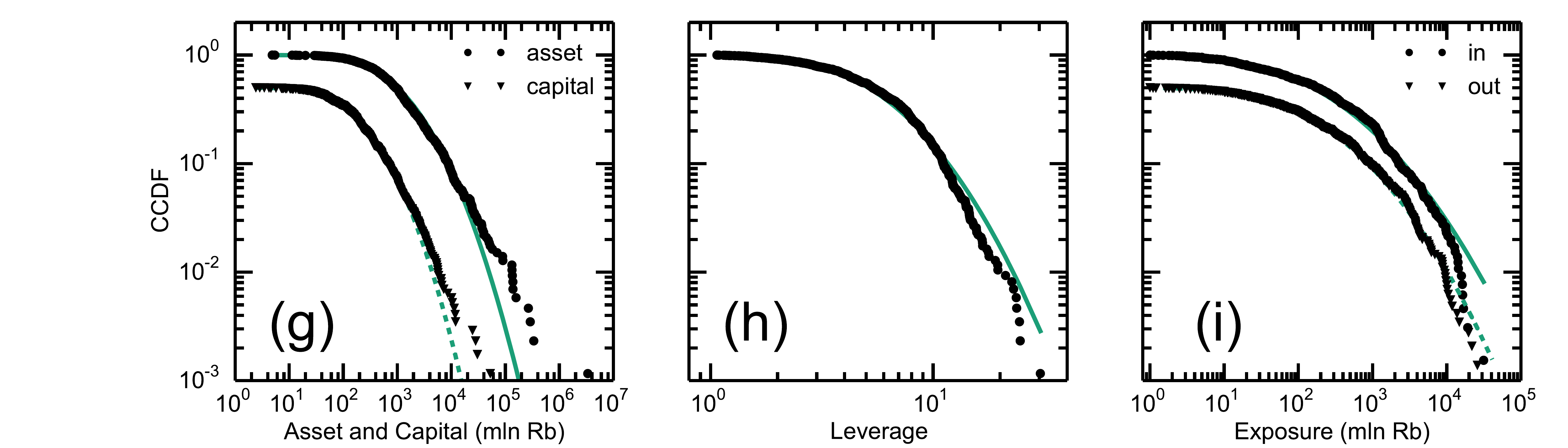}
\includegraphics[width=\textwidth, trim=0cm 0cm 0cm 0cm, clip=true]{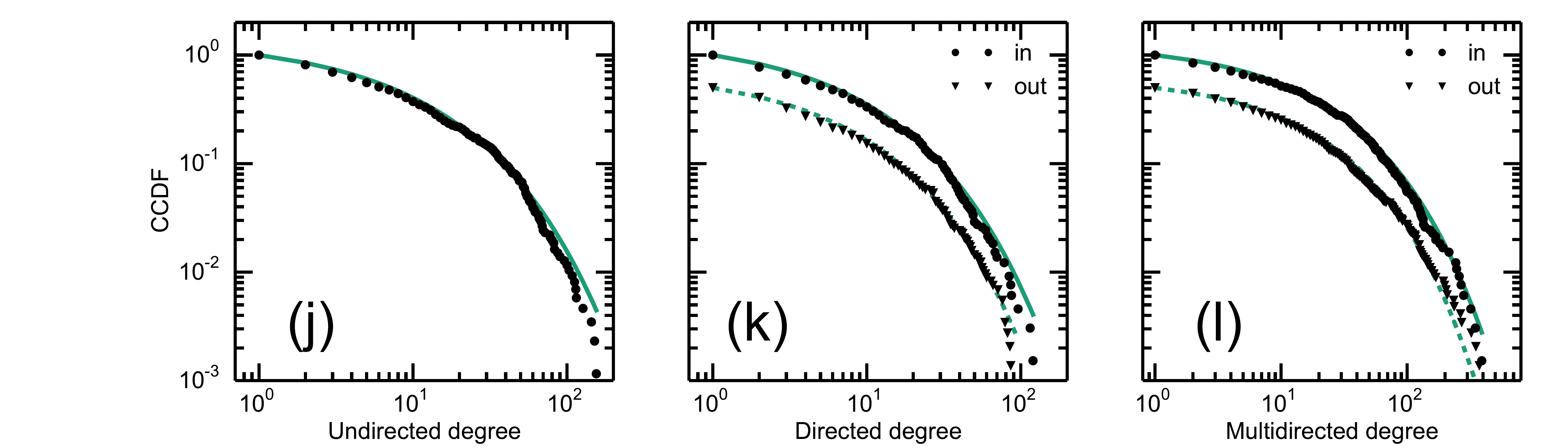}
\caption{(Color online) The CCDF for $10$ interbank network measures for the montly aggregated data in January 2004. Panels (a-f) display the tails parts of the data together with the PL (grey) and TPL (green) fits. Panels (g-l) contain the data over the entire range together with the "best" fit from Table~\ref{tab:ResultsTotalv2metJ1}. For illustrative reasons,  in those panels with two measures, the data of the second measure have been shifted vertically by multiplying them by a scaling factor of $0.5$.}
\label{fig:panel}
\end{center}
\end{figure*}

\section{Results}
\label{Sec:results}

The methodology used for analysing the out-exposures in the previous section, is used for each of the 12 complementary interbank network measures introduced in Sec.~\ref{Sec:data} and listed in Table~\ref{tab:ResultsTailsv2metJ1}. For a few combinations of network variables and time instances, we find that the $x_{min}$, which marks the lower boundary of the tail parts of the distributions and results from minimizing the quantity $Z$ of Eq.~(\ref{eq:KScriterion}), is close to the upper edge of the distribution. Under those circumstances the identified tail does not hold a sufficient amount of datapoints to perform a meaningful fit.  This is particularly common during the $1998$ interbank network collapse. For this reason, we only study the best-fit parameters from January 1999 onwards. 
 
A prototypical example of a data set for which the fit with the functions of Table~\ref{tab:DefinitionOfDistributions} is not fully satisfactory is shown in Fig.~\ref{fig:CPandLoan}.
The figure includes the loan size and counterparty-exposure measures for January 2004. Whereas the TPL is a good match for the tail parts of the data, the ``best" fit to the entire range of the data with a LN distribution clearly underestimates the probablility of events in the tail.  Fig.~\ref{fig:panel} displays for January 2004 the ``tail" and ``bulk+tail" parts of the 10 other interbank network measures considered. It is clear that the data can be reasonably well described by the adopted procedure and proposed theoretical distributions.  

Table~\ref{tab:ResultsTailsv2metJ1} reports our findings on the tail parts of the empirical distributions. The second column shows the percentage of data points assigned to the tail via the Kolmogorov-Smirnov criterion of Eq.~(\ref{eq:KScriterion}). This number averages to about $20$\% across the considered network measures but varies a lot. In particular, the loan size, counterparty exposure, and directed out-degree have fewer than $20$\% of the data-points in the tail. For most of the 70 months included in the analysis, the truncated power law outperforms, according to our methodology, the other four candidates for every measure. In Table~\ref{tab:ResultsTailsv2metJ1} the time-averaged values for the  $\alpha$ and $\lambda$ parameters are also reported. The power $\alpha$ is consistently of the order of two. It emerges from our studies that a fat tail is a characteristic feature of the measurable quantities of an interbank system. We notice that $\lambda$ is very volatile.

Upon evaluating the last column of Table~\ref{tab:ResultsTailsv2metJ1}, we find that although the truncated power law systematically provides the best fit, the other candidates are hardly significantly worse at the $1$ percent level. This observation is particularly true for the unweighted degree measures which are discrete. In fact, for the degree measures all directed and undirected versions lead to similar results. We find that  none of the five distribution candidates can be conclusively labeled as representing the best fit to the degree data. We note that the fraction of events located in the tail fluctuates a lot.

In order to evaluate the degree of sensitivity of our results to the amount of observations in the tail, we have made use of bootstrapping. For some prototypical measures we have drawn  $1000$ bootstrapped samples (random samples with replacement) of the real observations. The results of the analysis of the bootstrapped sample for the leverage, counterparty-exposure and directed out-degree are also included in Table~\ref{tab:ResultsTailsv2metJ1}. For the directed out-degree and the counterparty exposure, the fits to the real and bootstrapped data provide results which are compatible within the error bars. The leverage is the only studied network measure for which the best fit to the real data is provided by a PL. The PL, however, is the best fit for only 57\% of the studied months. For all studied months, the TPL emerges as a good alternate fit to the leverage data. Accordingly, it is not all that surprising that a TPL emerges as the best fit to the bootstrapped leverage data. All by all, we may conclude that despite the fact that the number of observations in the tails is rather volatile, the fits to the tails are rather robust.

The fact that the fraction of events in the tail is very volatile hints that a more integrated approach whereby the bulk and the tail parts of the data are simultaneously fitted with a non-Gaussian distribution may lead to a more robust description of the network features.  

Table~\ref{tab:ResultsTotalv2metJ1} summarizes our fits to the entire distributions of all 12 network measures over all 70 months using monthly-aggregated data. The asset, capital, leverage, loan sizes, as well as the counterparty-exposure and in-exposure distributions are described significantly best by a log-normal. The various degree distributions prefer the stretched exponential, which is on average not significantly better than the log-normal and/or the truncated power law. Similar results are obtained for the directed and undirected versions of the degree distributions. 
For those network measures for which the stretched exponential represents the best fit, we observe  that the parameter $\lambda$ is very volatile. It is worth noting that the stretched exponential has been put forward as a natural fat tail distribution for many physical and economic phenomena which cannot be satisfactorily described by a power law \cite{laherrere1998stretched}. In general, the parameters entering the fits to the tail parts are subject to larger fluctuations than those entering the distributions for the complete data set.

Our results compare well with some of the existing studies of interbank networks, but differ from others. Specifically, our findings are in line with Goddard et al.~\cite{goddard2014size} who study U.S. banks: the log-normal distribution fits the bulk part of the asset-size data well, while the power law does the same for the tail part. Our findings are not too different from those of Cont et al.~\cite{cont2013network} for the Brazilian interbank network: while our TPL fits to the tails of various degree distributions (see Table~\ref{tab:ResultsTailsv2metJ1}) deliver values of $\alpha$ between $1.7$ and $2.2$, Cont et al.~\cite{cont2013network} report power-law exponents between $2.2$ and $2.8$ (${\alpha}=2.54$ for total degree, ${\alpha}=2.46$ for in-degree, ${\alpha}=2.83$ for out-degree, and ${\alpha}=2.27$ for out-exposure size). Finally, in contrast to \cite{AustriaAnalysis} studying the Austrian interbank market we do not find that the power law provides a good fit to the entire loan-size distribution (\cite{AustriaAnalysis} report $\alpha=1.87$). Their power-law exponents reported for the tails of total and in-degree distributions (resp. $\alpha=2.0$ and $\alpha=3.11$) are comparable to the TPL values in Table~\ref{tab:ResultsTailsv2metJ1}, while their out-degree $\alpha=1.72$ differs substantially.

The most extensive study of interbank distributions was performed for the e-mid market~\cite{fricke2013distribution}. In line with this work, the authors consider both the complete and tail parts of the distributions and report results for daily and quarterly time aggregates. For the quarterly aggregates of the in-, out- and total degree distribution, a stretched exponential emerged as the best fit to the entire distribution. This confirms our findings. A log normal was reported as a best fit to the tail parts of the data. We also find that the LN can reasonably account for the tail parts. It is slightly outperformed by the truncated power law (which was not included in the set of possible distributions in the work of Ref.~\cite{fricke2013distribution}) and flags as not significantly worse in our table. For the quarterly aggregated number of transactions, a network feature comparable to the multidirected degree considered here, they also find similar results as for the regular degrees. 

Upon scrutiny of the time evolution of the fitted distributions and their parameters, we do not find any systematic changes in the distributions which emerge as best fit to the data, between the ``growth'' phase (1999-2002) and the ``mature'' phase (2002-2004). In line with the expectations, however, the extracted parameters are subject to smaller variations as the network grows and the number of nodes and edges increases.

\begin{table*}
\caption{Summary of the results of the fits to the monthly-aggregated
  data of all the network features considered in this work.  We list the distribution
  which provides the best fit to the full range of the data ("bulk"+"tail")), together with the percentage of months
  for which it comes out ``best'' according to the criterion of
  Eq.~(\ref{eq:sumofRs}). Columns 3 and 4 provide the average values
  of the parameters of the best-fit distribution. The last column list the distributions
  which are not significantly (at the 1 percent level) worse for more than
  half of the 70 months (January $1999$ - October $2004$) considered, together with the percentage of
  months for which this is the situation.}
\begin{tabular}{ l  c  c  c  c }
\hline \hline
	Measure 					& 	 Best fit (\%) & Parameter 1 & Parameter 2 & Alternative fit (\%) \\
	\hline

	Asset Size	 				&  	 LN (100) & $\mu = 5.45 \pm 0.63$ & $\sigma = 1.91 \pm 0.02$ & None\\
	Capital Size	 			&  	 LN (100) & $\mu = 4.64 \pm 0.53$ & $\sigma = 1.51 \pm 0.03$ & None\\
    Leverage 					& 	 LN (99) &	$\mu = 1.57 \pm 0.05$ &	$\sigma = 0.74 \pm 0.01$ & None \\
	Loan Size 					&  	 LN (100) & $\mu = 2.54 \pm 0.34$ & $\sigma = 1.27 \pm 0.08$ & None\\
	Counterparty-exposure		& 	 LN (99) & $\mu = 3.33 \pm 0.31$ & $\sigma = 1.56 \pm 0.13$ & None\\
	
	Undirected Degree 			& 	 SExp (54) & $\lambda = 0.29 \pm 0.20$ & $\beta = 0.55 \pm 0.06$ & LN (94)\\
	Directed In-Degree			& 	 SExp (66) & $\lambda = 0.41 \pm 0.27$ & $\beta = 0.46 \pm 0.07$ & TPL (91), LN (71) \\
	Directed Out-Degree 		& 	 SExp (83) & $\lambda = 0.43 \pm 0.39$ & $\beta = 0.60 \pm 0.05$ & TPL (51), LN (97) \\
	Multidirected In-Degree 	& 	 SExp (80) & $\lambda = 0.29 \pm 0.35$ & $\beta = 0.43 \pm 0.07$ & TPL (90) \\
	Multidirected Out-Degree	& 	 SExp (73) & $\lambda = 0.19 \pm 0.26$ & $\beta = 0.53 \pm 0.09$ & LN (80) \\

	In-Exposure		 			& 	 LN (83) & $\mu = 4.26 \pm 0.83$ & $\sigma = 2.37 \pm 0.09$ & None \\
	Out-Exposure	 			& 	 LN (80) & $\mu = 4.55 \pm 0.65$ & $\sigma = 2.13 \pm 0.10$ & SExp (57)\\
\hline \hline
\end{tabular}
\label{tab:ResultsTotalv2metJ1}
\end{table*}

As already mentioned, our data set covers two crises, the first in August 1998 and the second in the summer of 2004.  We would like to know if the best-fit candidates for the crisis periods are different than the ones for the normal operation periods.

As a consequence of the moratorium in August 1998, the interbank market collapsed in September, and then again in December 1998 (see Fig. \ref{fig:NodeTransactionNumber}). Due to this collapse, the network is too small and their are too few transactions to gather proper statistics. Fits to the data in this period are inconclusive for most of the measures, and hence are not included in the overall discussion.

During the second crisis period, the network does stay large enough to have proper statistics. From Fig.~\ref{fig:LR} it is clear that the stretched exponential distribution is significantly better than the log-normal from mid 2004 onwards. Thus, for the out-exposure (as well as the in-exposure), there is a difference between pre- and post-crisis best-fit candidates. For the other measures, however, we find that the preferred distributions are identical in "crisis" and in "normal" operation periods.

To end this section, we test the qualitative robustness of the obtained results using different time windows. This is particulary important in view of the fact that a recent study has shown that the interbank network properties depend on the aggregation period \cite{finger2013network}. We stress   that on average the number of banks active in one week is $77 \%$ of the monthly active ones. Accordingly, some banks are active in the interbank credit market during one week of a particular month. For the quarters and years this is respectively $114 \%$ and $132 \%$. This increase is mainly driven by new entrants to the market. Hence, banks tend to participate in the market at least once per month. To check if the preferred distribution is the same for different time aggregates, we repeat the monthly procedure for weekly, quarterly, and yearly aggregates. Table~\ref{tab:test} shows the ``best'' fit for each measure-aggregate pair and the percentage of the bins it is considered so. For the tail parts, we find that the TPL can be considered as the best overall fit candidate. We notice that in general the TPL becomes more preferred as the bin width increases. For each measure, the favoured fits to the entire distribution stay the same for different bin widths. We emphasize that the stability of the best-fit distribution with respect to variations in time do not imply that the parameters are not subject to time variations.

\begin{table*}
\caption{Robustness of the ``best'' fits for different aggregates. This table shows for each measure-aggregate pair the preferred fit candidate as well as the percentage of the bins it was considered so. This is done for fit of tail as well as total.}
\begin{tabular}{ l  c  c  c  c  c  c  c  c}
\hline \hline
								& 	\multicolumn{4}{c}{Tail} 				& 	\multicolumn{4}{c}{Bulk+Tail}\\
	Measure 					& 	Week	& 	Month	& 	Quarter & 	Year 	& 	Week 	& Month 		& 	Quarter & 	Year \\
	\hline
	
	Loan Size 					& TPL (70)	& TPL (75) 	& TPL (78) 	& TPL (80)	& LN (95)	&	LN (100)		& LN (100)	& LN (100) \\
	Counterparty-exposure		& TPL (82)	& TPL (80)	& TPL (83) 	& TPL (80)	& LN (94)	&	LN (99)		& LN (100)	& LN (100) \\
	
	Undirected Degree 			& TPL (94)	& TPL (100)	& TPL (90)	& TPL (100)	& SExp (66)	&	SExp (54)	& SExp (52) & SExp (80) \\
	Directed In-Degree			& TPL (93)	& TPL (94)	& TPL (90) 	& TPL (100)	& SExp (58) &	SExp (66)	& SExp (65) & SExp (60) \\
	Directed Out-Degree 		& TPL (83)	& TPL (89)	& TPL (95) 	& TPL (100)	& SExp (64) &	SExp (83)	& SExp (87) & SExp (100) \\
	Multidirected In-Degree 	& TPL (92)	& TPL (97)	& TPL (95) 	& TPL (100)	& SExp (78) &	SExp (80)	& SExp (87) & SExp (100)\\
	Multidirected Out-Degree	& TPL (79)	& TPL (93)	& TPL (90) 	& TPL (100)	& SExp (67)	&	SExp (73)	& SExp (87)	& SExp (100)\\

	In-Exposure		 			& TPL (94)	& TPL (97)	& TPL (100)	& TPL (100) & LN (74)	&	LN (83)		& LN (74) & LN (60) \\
	Out-Exposure 				& TPL (85)	& TPL (94)	& TPL (100) & TPL (100) & LN (83)	&	LN (80) 	& LN (57) & LN (80) \\

\hline \hline
\end{tabular}
\label{tab:test}
\end{table*}

\section{Conclusion}
\label{Sec:conclusion}
In this paper we use daily data on bilateral interbank exposures and monthly bank balance sheets to study network characteristics of the Russian interbank market over Aug 1998 - Oct 2004. Specifically, we examine the distributions of (un)directed (un)weighted degree, nodal attributes (bank assets, capital and capital-to-assets ratio) and edge weights (loan size and counterparty exposure). Using the methodology of ~\cite{ClausetPL} we set up a horse race between the different theoretical distributions to find one that fits the data best.

In line with the existing literature, we observe that all studied distributions are heavy tailed with the tail typically containing 20\% of the data. The tail is best described by a truncated power law, although the fit of other candidate distributions is only marginally worse. 
In most cases, separating the bulk and tail parts of the data turns out to be hard, and the proportion of observations assigned to the tail varies a lot. More stable fits to the data are obtained in an integrated approach that accounts for both the Gaussian and the non-Gaussian parts of the distributions. We find two distributions that fit the full range of the data best: the stretched exponential for measures related to unweighted degree and the log-normal for everything else. In case of the former, the log-normal performs only marginally worse. The power law distribution is rather ill suited to represent the full range of the studied characteristics of the Russian interbank market.

Our conclusions with regard to the best-fit distributions turn out to be robust to whether we aggregate the data over a week, month, quarter or year. Further, we find no qualitative difference between the ``growth'' and ``maturity'' phases of interbank market development and little difference between the ``crisis'' and ``non-crisis'' periods.

Our findings support the recent call ~\cite{ClausetPL,stumpf2012critical} for more rigorous statistical tests to detect power-law behavior in empirical data. While for most variables we find that the power law fits the tail of the distribution reasonably well, it is:
\begin{enumerate}
\item \emph{almost never the best} candidate to describe the tail;
\item \emph{typically not a good} candidate to describe the whole distribution. 
\end{enumerate}
Our findings echo those of Ref.~\cite{fricke2013distribution} who also find that the power law provides an inferior fit, compared to alternative distributions, to their overnight money market data coming from the e-MID trading platform. We thus tend to conclude that the evidence on power laws is not (yet) strong enough to warrant their widespread use in policy simulations.  From the study presented in this work we provide alternate distributions and corresponding parameters which can systematically and robustly capture the interbank network measures. We deem that those distributions represent a more realistic account of the interbank network structure than the widely used power laws. They could facilitate more realistic contagion modeling and provide more realistic estimates of interbank network measures.

\begin{acknowledgements}
This work is supported by the Research Foundation
Flanders (FWO-Flanders), the Research Foundation of Ghent University (BOF) and the Bank of Finland (BOFIT).
\end{acknowledgements}

\appendix

\bibliographystyle{unsrt}
\bibliography{bibliography}

\end{document}